\documentclass[10pt,nocopyrightspace,numbers,preprint]{sigplanconf}

\usepackage{bec}
\setboolean{showcomments}{false}
\setboolean{showjedi}{true}

\usepackage{rulelinks}

\newif\ifdraftfooter\draftfooterfalse

\usepackage{graphicx}
\usepackage{wrapfig}
\usepackage{amssymb}
\usepackage{amsmath}
\usepackage{amsthm}
\usepackage{semantic}
\usepackage{times}
\usepackage{mathpartir}
\usepackage{arydshln}
\usepackage{hyperref}
\usepackage{flushend}

\begin{document}
\newcommand{\ottdrule}[4][]{{\displaystyle\frac{\begin{array}{l}#2\end{array}}{#3}\quad\ottdrulename{#4}}}
\newcommand{\ottusedrule}[1]{\[#1\]}
\newcommand{\ottpremise}[1]{ #1 \\}
\newenvironment{ottdefnblock}[3][]{ \framebox{\mbox{#2}} \quad #3 \\[0pt]}{}
\newenvironment{ottfundefnblock}[3][]{ \framebox{\mbox{#2}} \quad #3 \\[0pt]\begin{displaymath}\begin{array}{l}}{\end{array}\end{displaymath}}
\newcommand{\ottfunclause}[2]{ #1 \equiv #2 \\}
\newcommand{\ottnt}[1]{\mathit{#1}}
\newcommand{\ottmv}[1]{\mathit{#1}}
\newcommand{\ottkw}[1]{\mathbf{#1}}
\newcommand{\ottsym}[1]{#1}
\newcommand{\ottcom}[1]{\text{#1}}
\newcommand{\ottdrulename}[1]{\textsc{#1}}
\newcommand{\ottcomplu}[5]{\overline{#1}^{\,#2\in #3 #4 #5}}
\newcommand{\ottcompu}[3]{\overline{#1}^{\,#2<#3}}
\newcommand{\ottcomp}[2]{\overline{#1}^{\,#2}}
\newcommand{\ottgrammartabular}[1]{\begin{supertabular}{llcllllll}#1\end{supertabular}}
\newcommand{\ottmetavartabular}[1]{\begin{supertabular}{ll}#1\end{supertabular}}
\newcommand{\ottrulehead}[3]{$#1$ & & $#2$ & & & \multicolumn{2}{l}{#3}}
\newcommand{\ottprodline}[6]{& & $#1$ & $#2$ & $#3 #4$ & $#5$ & $#6$}
\newcommand{\ottfirstprodline}[6]{\ottprodline{#1}{#2}{#3}{#4}{#5}{#6}}
\newcommand{\ottlongprodline}[2]{& & $#1$ & \multicolumn{4}{l}{$#2$}}
\newcommand{\ottfirstlongprodline}[2]{\ottlongprodline{#1}{#2}}
\newcommand{\ottbindspecprodline}[6]{\ottprodline{#1}{#2}{#3}{#4}{#5}{#6}}
\newcommand{\ottprodnewline}{\\}
\newcommand{\ottinterrule}{\\[5.0mm]}
\newcommand{\ottafterlastrule}{\\}
\newcommand{\ottmetavars}{
\ottmetavartabular{
 $ \ottmv{var} ,\, x ,\, y ,\, z ,\, f ,\, g $ & \ottcom{variables} \\
 $ \ottmv{n} ,\, \ottmv{s} ,\, \ottmv{numeral} $ & \ottcom{literal number} \\
 $ i ,\, j $ &  \\
 $ \ell $ & \ottcom{store locations} \\
}}

\newcommand{\ottJ}{
\ottrulehead{\ottnt{J}}{::=}{\ottcom{Custom judgement forms; for TeX}}\ottprodnewline
\ottfirstprodline{|}{ \sigma  \longrightarrow  \sigma' }{}{}{}{}\ottprodnewline
\ottprodline{|}{ \sigma  & \longrightarrow &  \sigma' }{}{}{}{}\ottprodnewline
\ottprodline{|}{\Gamma  \vdash  \ottnt{v}  \ottsym{<~~}  \ottnt{A}}{}{}{}{}\ottprodnewline
\ottprodline{|}{ \rho ( \ottnt{v} )\,{=}\, \ottnt{v'} }{}{}{}{}\ottprodnewline
\ottprodline{|}{ \ell  \not\in  \mu }{}{}{}{}\ottprodnewline
\ottprodline{|}{\Gamma  \vdash  \dot{e}  \ottsym{:}  \ottnt{C}}{}{}{}{}\ottprodnewline
\ottprodline{|}{\Gamma  \vdash  \dot{v}  \ottsym{:}  \ottnt{A}}{}{}{}{}\ottprodnewline
\ottprodline{|}{\Gamma  \vdash  \ottnt{e}  \ottsym{:}  \ottnt{C}}{}{}{}{}\ottprodnewline
\ottprodline{|}{\Gamma  \vdash  \ottnt{v}  \ottsym{:}  \ottnt{A}}{}{}{}{}\ottprodnewline
\ottprodline{|}{\Gamma  \vdash  \ottnt{e}  \Rightarrow  \ottnt{C}}{}{}{}{}\ottprodnewline
\ottprodline{|}{\Gamma  \vdash  \ottnt{e}  \Leftarrow  \ottnt{C}}{}{}{}{}\ottprodnewline
\ottprodline{|}{\Gamma  \vdash  \ottnt{v}  \Leftarrow  \ottnt{A}}{}{}{}{}\ottprodnewline
\ottprodline{|}{\Gamma  \vdash  \ottnt{v}  \Rightarrow  \ottnt{A}}{}{}{}{}\ottprodnewline
\ottprodline{|}{\Gamma  \vdash  \dot{e}  \Rightarrow  \ottnt{C}}{}{}{}{}\ottprodnewline
\ottprodline{|}{\Gamma  \vdash  \dot{e}  \Leftarrow  \ottnt{C}}{}{}{}{}\ottprodnewline
\ottprodline{|}{\Gamma  \vdash  \dot{v}  \Leftarrow  \ottnt{A}}{}{}{}{}\ottprodnewline
\ottprodline{|}{\Gamma  \vdash  \dot{v}  \Rightarrow  \ottnt{A}}{}{}{}{}\ottprodnewline
\ottprodline{|}{\Gamma_{{\mathrm{1}}}  \vdash  \rho  \Rightarrow  \Gamma_{{\mathrm{2}}}}{}{}{}{}\ottprodnewline
\ottprodline{|}{\Gamma  \vdash  \dot{e}  \Rightarrow  \ottnt{C}}{}{}{}{}\ottprodnewline
\ottprodline{|}{\Gamma  \vdash  \kappa  \Leftarrow  \ottnt{C}}{}{}{}{}\ottprodnewline
\ottprodline{|}{\sigma \, {\textsf{ok} }}{}{}{}{}\ottprodnewline
\ottprodline{|}{ e_\textsf{m}  ([\![  \sigma  ]\!]) \Downarrow_{\textsf{meta} } [\![  \sigma'  ]\!] }{}{}{}{}\ottprodnewline
\ottprodline{|}{ \ottnt{A}  \approx_{\textsf{?} }  \ottnt{B} }{}{}{}{}\ottprodnewline
\ottprodline{|}{ \ottnt{C}  \approx_{\textsf{?} }  \ottnt{D} }{}{}{}{}\ottprodnewline
\ottprodline{|}{\Delta_{{\mathrm{1}}}  \ottsym{=}  \Delta_{{\mathrm{2}}}}{}{}{}{}\ottprodnewline
\ottprodline{|}{ \ottnt{A}  \notin  \ottnt{B} }{}{}{}{}}

\newcommand{\ottA}{
\ottrulehead{\ottnt{A}  ,\ \ottnt{B}}{::=}{\ottcom{Value Types}}\ottprodnewline
\ottfirstprodline{|}{\textsf{?}}{}{}{}{}\ottprodnewline
\ottprodline{|}{\ottkw{Num}}{}{}{}{}\ottprodnewline
\ottprodline{|}{\ottkw{Str}}{}{}{}{}\ottprodnewline
\ottprodline{|}{\ottkw{Dict} \, \Delta}{}{}{}{}\ottprodnewline
\ottprodline{|}{\ottkw{Ref} \, \ottnt{A}}{}{}{}{}\ottprodnewline
\ottprodline{|}{\ottkw{U} \, \ottnt{C}}{}{}{}{}\ottprodnewline
\ottprodline{|}{\ottsym{(}  \ottnt{A}  \ottsym{)}} {\textsf{S}}{}{}{}\ottprodnewline
\ottprodline{|}{\ottsym{1}} {\textsf{S}}{}{}{}\ottprodnewline
\ottprodline{|}{\ottsym{2}} {\textsf{S}}{}{}{}\ottprodnewline
\ottprodline{|}{\ottkw{Bool}} {\textsf{S}}{}{}{}\ottprodnewline
\ottprodline{|}{\ottkw{Db} \, \ottnt{A}}{}{}{}{}}

\newcommand{\ottDelta}{
\ottrulehead{\Delta}{::=}{\ottcom{Dictionary types: Internal structure}}\ottprodnewline
\ottfirstprodline{|}{\varepsilon}{}{}{}{}\ottprodnewline
\ottprodline{|}{ \Delta ,  \ottnt{v} \mapsto \ottnt{A} }{}{}{}{}\ottprodnewline
\ottprodline{|}{\ottsym{(}  \Delta  \ottsym{)}} {\textsf{S}}{}{}{}}

\newcommand{\ottDeltas}{
\ottrulehead{\Delta}{::=}{}\ottprodnewline
\ottfirstprodline{|}{\Delta}{}{}{}{}\ottprodnewline
\ottprodline{|}{ \Delta_{{\mathrm{1}}}  ,  \Delta_{{\mathrm{2}}} }{}{}{}{}\ottprodnewline
\ottprodline{|}{\ottsym{(}  \Delta  \ottsym{)}} {\textsf{S}}{}{}{}}

\newcommand{\ottC}{
\ottrulehead{\ottnt{C}  ,\ \ottnt{D}}{::=}{\ottcom{Computation Types}}\ottprodnewline
\ottfirstprodline{|}{\ottnt{A}  \rightarrow  \ottnt{C}}{}{}{}{}\ottprodnewline
\ottprodline{|}{\ottkw{F} \, \ottnt{A}}{}{}{}{}\ottprodnewline
\ottprodline{|}{\ottsym{(}  \ottnt{C}  \ottsym{)}} {\textsf{S}}{}{}{}}

\newcommand{\ottae}{
\ottrulehead{a^\textsf{e}}{::=}{\ottcom{Annotations for expressions}}\ottprodnewline
\ottfirstprodline{|}{\ottnt{C}}{}{}{}{}\ottprodnewline
\ottprodline{|}{ \_ }{}{}{}{}}

\newcommand{\ottav}{
\ottrulehead{a^\textsf{v}}{::=}{\ottcom{Anotations for values}}\ottprodnewline
\ottfirstprodline{|}{\ottnt{A}}{}{}{}{}\ottprodnewline
\ottprodline{|}{ \_ }{}{}{}{}}

\newcommand{\otte}{
\ottrulehead{\ottnt{e}}{::=}{\ottcom{Expressions: Annotated pre-expressions}}\ottprodnewline
\ottfirstprodline{|}{ \dot{e} ~\texttt{@:}~ a^\textsf{e} }{}{}{}{}\ottprodnewline
\ottprodline{|}{\ottsym{(}  \ottnt{e}  \ottsym{)}} {\textsf{S}}{}{}{}}

\newcommand{\ottme}{
\ottrulehead{e_\textsf{m}}{::=}{\ottcom{Meta-Level Expressions}}}

\newcommand{\ottpe}{
\ottrulehead{\dot{e}}{::=}{\ottcom{Pre-Expressions}}\ottprodnewline
\ottfirstprodline{|}{\ottnt{e} \, \ottnt{v}}{}{}{}{}\ottprodnewline
\ottprodline{|}{ \lambda  x .  \ottnt{e} }{}{}{}{}\ottprodnewline
\ottprodline{|}{ \ottkw{let} \; x \,{\texttt{=} }\, \ottnt{e_{{\mathrm{1}}}} \, \ottkw{in} \, \ottnt{e_{{\mathrm{2}}}} }{}{}{}{}\ottprodnewline
\ottprodline{|}{\ottkw{ret} \, \ottnt{v}}{}{}{}{}\ottprodnewline
\ottprodline{|}{\ottkw{force} \, \ottnt{v}}{}{}{}{}\ottprodnewline
\ottprodline{|}{\ottkw{ref} \, \ottnt{v}}{}{}{}{}\ottprodnewline
\ottprodline{|}{\ottkw{set} \, \ottnt{v_{{\mathrm{1}}}} \, \ottnt{v_{{\mathrm{2}}}}}{}{}{}{}\ottprodnewline
\ottprodline{|}{\ottkw{get} \, \ottnt{v}}{}{}{}{}\ottprodnewline
\ottprodline{|}{\ottkw{ext} \, \ottnt{v_{{\mathrm{1}}}} \, \ottnt{v_{{\mathrm{2}}}} \, \ottnt{v_{{\mathrm{3}}}}}{}{}{}{}\ottprodnewline
\ottprodline{|}{ \ottnt{v_{{\mathrm{1}}}} \left[  \ottnt{v_{{\mathrm{2}}}}  \right]_{ a^{\textsf{op} } } }{}{}{}{}\ottprodnewline
\ottprodline{|}{\ottnt{e}  \mathrel{ \texttt{?:} }  a^\textsf{e}}{}{}{}{}\ottprodnewline
\ottprodline{|}{\ottkw{rcc} \, e_\textsf{m} \, \ottnt{e}}{}{}{}{}\ottprodnewline
\ottprodline{|}{ \texttt{openDb}_{ a^{\textsf{op} } }~ \ottnt{v} }{}{}{}{}\ottprodnewline
\ottprodline{|}{ \texttt{filterDb}_{ a^{\textsf{op} } }~ \ottnt{v_{{\mathrm{1}}}} ~ \ottnt{v_{{\mathrm{2}}}} }{}{}{}{}\ottprodnewline
\ottprodline{|}{ \texttt{joinDb}_{ a^{\textsf{op} } }~ \ottnt{v_{{\mathrm{1}}}} ~ \ottnt{v_{{\mathrm{2}}}} ~ \ottnt{v_{{\mathrm{3}}}} ~ \ottnt{v_{{\mathrm{4}}}} }{}{}{}{}\ottprodnewline
\ottprodline{|}{\ottsym{(}  \ottnt{e}  \ottsym{)}} {\textsf{S}}{}{}{}}

\newcommand{\ottoa}{
\ottrulehead{a^{\textsf{op} }}{::=}{\ottcom{Operation annotation}}\ottprodnewline
\ottfirstprodline{|}{ \texttt{?} }{}{}{}{\ottcom{Top modality; some info may be missing}}\ottprodnewline
\ottprodline{|}{ \texttt{!} }{}{}{}{\ottcom{Ground modality; no info is missing}}}

\newcommand{\ottenv}{
\ottrulehead{\rho}{::=}{}\ottprodnewline
\ottfirstprodline{|}{\varepsilon}{}{}{}{}\ottprodnewline
\ottprodline{|}{ \rho  ,  x \mapsto \ottnt{v} }{}{}{}{}}

\newcommand{\ottstack}{
\ottrulehead{\kappa}{::=}{}\ottprodnewline
\ottfirstprodline{|}{\textsf{halt}}{}{}{}{}\ottprodnewline
\ottprodline{|}{\kappa  ::  \ottnt{v}}{}{}{}{}\ottprodnewline
\ottprodline{|}{\kappa  ::  \ottsym{(}  \rho  \ottsym{,}  x  \ottsym{.}  \ottnt{e_{{\mathrm{2}}}}  \ottsym{)}}{}{}{}{}}

\newcommand{\ottstore}{
\ottrulehead{\mu}{::=}{}\ottprodnewline
\ottfirstprodline{|}{\varepsilon}{}{}{}{}\ottprodnewline
\ottprodline{|}{ \mu  ,  \ell \mapsto \ottnt{v} }{}{}{}{}}

\newcommand{\ottstate}{
\ottrulehead{\sigma}{::=}{}\ottprodnewline
\ottfirstprodline{|}{ \left<  \mu ;  \kappa ;  \rho ;  \dot{e}  \right> }{}{}{}{}\ottprodnewline
\ottprodline{|}{ \left<  \mu ;  \kappa ;  \rho  \right. & \left.  \dot{e}  \right> }{}{}{}{}}

\newcommand{\ottv}{
\ottrulehead{\ottnt{v}}{::=}{\ottcom{Values: Annotated Pre-Values}}\ottprodnewline
\ottfirstprodline{|}{ \dot{v} ~\texttt{@:}~ a^\textsf{v} }{}{}{}{}\ottprodnewline
\ottprodline{|}{\ottsym{(}  \ottnt{v}  \ottsym{)}} {\textsf{S}}{}{}{}\ottprodnewline
\ottprodline{|}{ \rho ( \ottnt{v} ) } {\textsf{M}}{}{}{}}

\newcommand{\ottpv}{
\ottrulehead{\dot{v}}{::=}{\ottcom{Pre-Values}}\ottprodnewline
\ottfirstprodline{|}{\ottkw{othunk} \, \ottnt{e}}{}{}{}{}\ottprodnewline
\ottprodline{|}{\ottkw{thunk} \, \rho \, \ottnt{e}}{}{}{}{}\ottprodnewline
\ottprodline{|}{\ottkw{dict} \, \delta}{}{}{}{}\ottprodnewline
\ottprodline{|}{\ottkw{num} \, \ottmv{n}}{}{}{}{}\ottprodnewline
\ottprodline{|}{\ottkw{str} \, \ottmv{s}}{}{}{}{}\ottprodnewline
\ottprodline{|}{\ottkw{loc} \, \ell}{}{}{}{}\ottprodnewline
\ottprodline{|}{\ottkw{bool} \, \ottnt{b}}{}{}{}{}\ottprodnewline
\ottprodline{|}{x}{}{}{}{}\ottprodnewline
\ottprodline{|}{\ottsym{()}} {\textsf{S}}{}{}{}\ottprodnewline
\ottprodline{|}{\ottsym{(}  \dot{v}  \ottsym{)}} {\textsf{S}}{}{}{}}

\newcommand{\ottb}{
\ottrulehead{\ottnt{b}}{::=}{\ottcom{Bits}}}

\newcommand{\ottdelta}{
\ottrulehead{\delta}{::=}{\ottcom{Dictionary values: Internal structure}}\ottprodnewline
\ottfirstprodline{|}{\varepsilon}{}{}{}{}\ottprodnewline
\ottprodline{|}{ \delta  ,  \ottnt{v_{{\mathrm{1}}}}  \mapsto  \ottnt{v_{{\mathrm{2}}}} }{}{}{}{}\ottprodnewline
\ottprodline{|}{\ottsym{(}  \delta  \ottsym{)}}{}{}{}{}}

\newcommand{\ottCtx}{
\ottrulehead{\Gamma}{::=}{}\ottprodnewline
\ottfirstprodline{|}{\varepsilon}{}{}{}{}\ottprodnewline
\ottprodline{|}{\Gamma  \ottsym{,}  x  \ottsym{:}  \ottnt{A}}{}{}{}{}\ottprodnewline
\ottprodline{|}{ \left| \mu \right| }{}{}{}{}\ottprodnewline
\ottprodline{|}{\Gamma_{{\mathrm{1}}}  \ottsym{,}  \Gamma_{{\mathrm{2}}}} {\textsf{S}}{}{}{}}

\newcommand{\ottterminals}{
\ottrulehead{\ottnt{terminals}}{::=}{}\ottprodnewline
\ottfirstprodline{|}{ \Rightarrow }{}{}{}{}\ottprodnewline
\ottprodline{|}{ \Leftarrow }{}{}{}{}\ottprodnewline
\ottprodline{|}{ \mathrel{ \texttt{?:} } }{}{}{}{}\ottprodnewline
\ottprodline{|}{ :: }{}{}{}{}\ottprodnewline
\ottprodline{|}{ \Lambda }{}{}{}{}\ottprodnewline
\ottprodline{|}{ \forall }{}{}{}{}\ottprodnewline
\ottprodline{|}{ \rhd }{}{}{}{}\ottprodnewline
\ottprodline{|}{ (\!| }{}{}{}{}\ottprodnewline
\ottprodline{|}{ |\!) }{}{}{}{}\ottprodnewline
\ottprodline{|}{ {\bullet} }{}{}{}{}\ottprodnewline
\ottprodline{|}{ {\circ} }{}{}{}{}\ottprodnewline
\ottprodline{|}{ \circ }{}{}{}{}\ottprodnewline
\ottprodline{|}{ \ne }{}{}{}{}\ottprodnewline
\ottprodline{|}{ \leadsto }{}{}{}{}\ottprodnewline
\ottprodline{|}{ \leadsto }{}{}{}{}\ottprodnewline
\ottprodline{|}{ \longrightarrow }{}{}{}{}\ottprodnewline
\ottprodline{|}{ \bigtriangleup }{}{}{}{}\ottprodnewline
\ottprodline{|}{ \bigtriangledown }{}{}{}{}\ottprodnewline
\ottprodline{|}{ \curvearrowright }{}{}{}{}\ottprodnewline
\ottprodline{|}{ \curvearrowright_{\textsf{alg} } }{}{}{}{}\ottprodnewline
\ottprodline{|}{ \curvearrowright_{\textsf{prop} } }{}{}{}{}\ottprodnewline
\ottprodline{|}{ \Downarrow }{}{}{}{}\ottprodnewline
\ottprodline{|}{ \downarrow }{}{}{}{}\ottprodnewline
\ottprodline{|}{ \approx }{}{}{}{}\ottprodnewline
\ottprodline{|}{ \square }{}{}{}{}\ottprodnewline
\ottprodline{|}{ \mathrel{\texttt{>\!>=} } }{}{}{}{}\ottprodnewline
\ottprodline{|}{ \textsf{halt} }{}{}{}{}\ottprodnewline
\ottprodline{|}{ \textsf{prop} }{}{}{}{}\ottprodnewline
\ottprodline{|}{ \circlearrowleft }{}{}{}{}\ottprodnewline
\ottprodline{|}{ \lambda }{}{}{}{}\ottprodnewline
\ottprodline{|}{ \alpha }{}{}{}{}\ottprodnewline
\ottprodline{|}{ \beta }{}{}{}{}\ottprodnewline
\ottprodline{|}{ \bot }{}{}{}{}\ottprodnewline
\ottprodline{|}{ \top }{}{}{}{}\ottprodnewline
\ottprodline{|}{ \rightarrow }{}{}{}{}\ottprodnewline
\ottprodline{|}{ \Rightarrow }{}{}{}{}\ottprodnewline
\ottprodline{|}{ \Leftarrow }{}{}{}{}\ottprodnewline
\ottprodline{|}{ \mathrel{ {\Leftarrow}_{\mathsf{V} } } }{}{}{}{}\ottprodnewline
\ottprodline{|}{ \mathrel{ {\Leftarrow}_{\mathsf{E} } } }{}{}{}{}\ottprodnewline
\ottprodline{|}{ \multimap }{}{}{}{}\ottprodnewline
\ottprodline{|}{ \leftarrow }{}{}{}{}\ottprodnewline
\ottprodline{|}{ {\mapsto} }{}{}{}{}\ottprodnewline
\ottprodline{|}{ \leadsto }{}{}{}{}\ottprodnewline
\ottprodline{|}{ \hookrightarrow }{}{}{}{}\ottprodnewline
\ottprodline{|}{ \left<\right. }{}{}{}{}\ottprodnewline
\ottprodline{|}{ \left.\right> }{}{}{}{}\ottprodnewline
\ottprodline{|}{ \cdot }{}{}{}{}\ottprodnewline
\ottprodline{|}{ [] }{}{}{}{}\ottprodnewline
\ottprodline{|}{ \wedge }{}{}{}{}\ottprodnewline
\ottprodline{|}{ \vee }{}{}{}{}\ottprodnewline
\ottprodline{|}{ \subseteq }{}{}{}{}\ottprodnewline
\ottprodline{|}{ \dashv }{}{}{}{}\ottprodnewline
\ottprodline{|}{ \vdash }{}{}{}{}\ottprodnewline
\ottprodline{|}{ \Vdash }{}{}{}{}\ottprodnewline
\ottprodline{|}{ \not\vdash }{}{}{}{}\ottprodnewline
\ottprodline{|}{ \vDash }{}{}{}{}\ottprodnewline
\ottprodline{|}{ \Delta }{}{}{}{}\ottprodnewline
\ottprodline{|}{ \textsf{?} }{}{}{}{}\ottprodnewline
\ottprodline{|}{ \cup }{}{}{}{}\ottprodnewline
\ottprodline{|}{ \in }{}{}{}{}\ottprodnewline
\ottprodline{|}{ \notin }{}{}{}{}\ottprodnewline
\ottprodline{|}{ {\mathrel{@} } }{}{}{}{}\ottprodnewline
\ottprodline{|}{ \varepsilon }{}{}{}{}\ottprodnewline
\ottprodline{|}{ {\textsf{ok} } }{}{}{}{}\ottprodnewline
\ottprodline{|}{ {\textsf{wf} } }{}{}{}{}}

\newcommand{\ottformula}{
\ottrulehead{\ottnt{formula}}{::=}{}\ottprodnewline
\ottfirstprodline{|}{\ottnt{judgement}}{}{}{}{}}

\newcommand{\ottJtyp}{
\ottrulehead{\ottnt{Jtyp}}{::=}{}\ottprodnewline
\ottfirstprodline{|}{ \textsf{dom}( \rho ) \supseteq \textsf{FV}(e) }{}{}{}{}\ottprodnewline
\ottprodline{|}{ \Gamma   \vdash   \ottnt{v}  ~~\textsf{closed} }{}{}{}{\ottcom{Invariants for closed values}}\ottprodnewline
\ottprodline{|}{ \Gamma   \vdash   \dot{v}  ~~\textsf{closed} }{}{}{}{\ottcom{Invariants for closed values}}\ottprodnewline
\ottprodline{|}{\Gamma  \vdash  \ottnt{e}  \ottsym{:}  \ottnt{C}}{}{}{}{\ottcom{Expression Typing}}\ottprodnewline
\ottprodline{|}{\Gamma  \vdash  \dot{e}  \ottsym{:}  \ottnt{C}}{}{}{}{\ottcom{Computation Typing}}\ottprodnewline
\ottprodline{|}{ \ottnt{C}  \approx  \ottnt{D} }{}{}{}{}\ottprodnewline
\ottprodline{|}{ \ottnt{A}  \approx  \ottnt{B} }{}{}{}{}\ottprodnewline
\ottprodline{|}{ \Gamma   \vdash   \ottnt{v}  \Leftarrow  \ottnt{A} }{}{}{}{\ottcom{Value Checking}}\ottprodnewline
\ottprodline{|}{\Gamma  \vdash  \ottnt{v}  \ottsym{:}  \ottnt{A}}{}{}{}{\ottcom{Value Typing}}\ottprodnewline
\ottprodline{|}{\Gamma  \vdash  \dot{v}  \ottsym{:}  \ottnt{A}}{}{}{}{\ottcom{Value Typing}}\ottprodnewline
\ottprodline{|}{\Gamma  \vdash  \rho  \leadsto  \Gamma'}{}{}{}{\ottcom{Environment Typing}}\ottprodnewline
\ottprodline{|}{ \Gamma   \vdash   \kappa  \Leftarrow  \ottnt{C} }{}{}{}{\ottcom{Stack Typing}}\ottprodnewline
\ottprodline{|}{\sigma \, {\textsf{ok} }}{}{}{}{}\ottprodnewline
\ottprodline{|}{\sigma \, \ottkw{final}}{}{}{}{}\ottprodnewline
\ottprodline{|}{ \rho ( x ) =  \ottnt{v} }{}{}{}{}\ottprodnewline
\ottprodline{|}{ \rho ( \delta ) \leadsto  \delta' }{}{}{}{}\ottprodnewline
\ottprodline{|}{ \rho ( \ottnt{v} ) \leadsto  \ottnt{v'} }{}{}{}{}\ottprodnewline
\ottprodline{|}{ \ell  \not\in  \mu }{}{}{}{}\ottprodnewline
\ottprodline{|}{ [\![  \sigma  ]\!] \vdash  e_\textsf{m}  \Downarrow_{\textsf{rcc} } state' }{}{}{}{}\ottprodnewline
\ottprodline{|}{\sigma  \longrightarrow  \sigma'}{}{}{}{}}

\newcommand{\ottjudgement}{
\ottrulehead{\ottnt{judgement}}{::=}{}\ottprodnewline
\ottfirstprodline{|}{\ottnt{Jtyp}}{}{}{}{}}

\newcommand{\ottuserXXsyntax}{
\ottrulehead{\ottnt{user\_syntax}}{::=}{}\ottprodnewline
\ottfirstprodline{|}{\ottmv{var}}{}{}{}{}\ottprodnewline
\ottprodline{|}{\ottmv{n}}{}{}{}{}\ottprodnewline
\ottprodline{|}{i}{}{}{}{}\ottprodnewline
\ottprodline{|}{\ell}{}{}{}{}\ottprodnewline
\ottprodline{|}{\ottnt{J}}{}{}{}{}\ottprodnewline
\ottprodline{|}{\ottnt{A}}{}{}{}{}\ottprodnewline
\ottprodline{|}{\Delta}{}{}{}{}\ottprodnewline
\ottprodline{|}{\Delta}{}{}{}{}\ottprodnewline
\ottprodline{|}{\ottnt{C}}{}{}{}{}\ottprodnewline
\ottprodline{|}{a^\textsf{e}}{}{}{}{}\ottprodnewline
\ottprodline{|}{a^\textsf{v}}{}{}{}{}\ottprodnewline
\ottprodline{|}{\ottnt{e}}{}{}{}{}\ottprodnewline
\ottprodline{|}{e_\textsf{m}}{}{}{}{}\ottprodnewline
\ottprodline{|}{\dot{e}}{}{}{}{}\ottprodnewline
\ottprodline{|}{a^{\textsf{op} }}{}{}{}{}\ottprodnewline
\ottprodline{|}{\rho}{}{}{}{}\ottprodnewline
\ottprodline{|}{\kappa}{}{}{}{}\ottprodnewline
\ottprodline{|}{\mu}{}{}{}{}\ottprodnewline
\ottprodline{|}{\sigma}{}{}{}{}\ottprodnewline
\ottprodline{|}{\ottnt{v}}{}{}{}{}\ottprodnewline
\ottprodline{|}{\dot{v}}{}{}{}{}\ottprodnewline
\ottprodline{|}{\ottnt{b}}{}{}{}{}\ottprodnewline
\ottprodline{|}{\delta}{}{}{}{}\ottprodnewline
\ottprodline{|}{\Gamma}{}{}{}{}\ottprodnewline
\ottprodline{|}{\ottnt{terminals}}{}{}{}{}}

\newcommand{\ottgrammar}{\ottgrammartabular{
\ottJ\ottinterrule
\ottA\ottinterrule
\ottDelta\ottinterrule
\ottDeltas\ottinterrule
\ottC\ottinterrule
\ottae\ottinterrule
\ottav\ottinterrule
\otte\ottinterrule
\ottme\ottinterrule
\ottpe\ottinterrule
\ottoa\ottinterrule
\ottenv\ottinterrule
\ottstack\ottinterrule
\ottstore\ottinterrule
\ottstate\ottinterrule
\ottv\ottinterrule
\ottpv\ottinterrule
\ottb\ottinterrule
\ottdelta\ottinterrule
\ottCtx\ottinterrule
\ottterminals\ottinterrule
\ottformula\ottinterrule
\ottJtyp\ottinterrule
\ottjudgement\ottinterrule
\ottuserXXsyntax\ottafterlastrule
}}


\newcommand{\ottdefndomXXenvXXcontainsXXfvXXe}[1]{\begin{ottdefnblock}[#1]{$ \textsf{dom}( \rho ) \supseteq \textsf{FV}(e) $}{}
\end{ottdefnblock}}


\newcommand{\ottdefnvXXclosed}[1]{\begin{ottdefnblock}[#1]{$ \Gamma   \vdash   \ottnt{v}  ~~\textsf{closed} $}{\ottcom{Invariants for closed values}}
\end{ottdefnblock}}

\newcommand{\ottdruleloc}[1]{\ottdrule[#1]{%
}{
  \left|  \mu  ,  \ell \mapsto \ottnt{v}  \right|    \vdash   \ottkw{loc} \, \ell  ~~\textsf{closed} }{%
{\ottdrulename{loc}}{}%
}}

\newcommand{\ottdrulestr}[1]{\ottdrule[#1]{%
}{
  \left| \mu \right|    \vdash   \ottkw{str} \, \ottmv{s}  ~~\textsf{closed} }{%
{\ottdrulename{str}}{}%
}}

\newcommand{\ottdrulenum}[1]{\ottdrule[#1]{%
}{
  \left| \mu \right|    \vdash   \ottkw{num} \, \ottmv{n}  ~~\textsf{closed} }{%
{\ottdrulename{num}}{}%
}}

\newcommand{\ottdruledictXXemp}[1]{\ottdrule[#1]{%
}{
  \left| \mu \right|    \vdash   \ottkw{dict} \, \varepsilon  ~~\textsf{closed} }{%
{\ottdrulename{dict\_emp}}{}%
}}

\newcommand{\ottdruledictXXcons}[1]{\ottdrule[#1]{%
\ottpremise{  \left| \mu \right|    \vdash   \ottkw{dict} \, \delta  ~~\textsf{closed} }%
\ottpremise{  \left| \mu \right|    \vdash   \ottnt{v_{{\mathrm{1}}}}  ~~\textsf{closed} }%
\ottpremise{  \left| \mu \right|    \vdash   \ottnt{v_{{\mathrm{2}}}}  ~~\textsf{closed} }%
}{
  \left| \mu \right|    \vdash   \ottkw{dict} \, \ottsym{(}   \delta  ,  \ottnt{v_{{\mathrm{1}}}}  \mapsto  \ottnt{v_{{\mathrm{2}}}}   \ottsym{)}  ~~\textsf{closed} }{%
{\ottdrulename{dict\_cons}}{}%
}}

\newcommand{\ottdrulethunk}[1]{\ottdrule[#1]{%
\ottpremise{ \textsf{dom}( \rho ) \supseteq \textsf{FV}(e) }%
}{
  \left| \mu \right|    \vdash   \ottkw{thunk} \, \rho \, \ottnt{e}  ~~\textsf{closed} }{%
{\ottdrulename{thunk}}{}%
}}

\newcommand{\ottdefnpvXXclosed}[1]{\begin{ottdefnblock}[#1]{$ \Gamma   \vdash   \dot{v}  ~~\textsf{closed} $}{\ottcom{Invariants for closed values}}
\ottusedrule{\ottdruleloc{}}
\ottusedrule{\ottdrulestr{}}
\ottusedrule{\ottdrulenum{}}
\ottusedrule{\ottdruledictXXemp{}}
\ottusedrule{\ottdruledictXXcons{}}
\ottusedrule{\ottdrulethunk{}}
\end{ottdefnblock}}

\newcommand{\ottdruletye}[1]{\ottdrule[#1]{%
\ottpremise{\Gamma  \vdash  \dot{e}  \ottsym{:}  \ottnt{C}}%
}{
\Gamma  \vdash  \ottsym{(}   \dot{e} ~\texttt{@:}~ a^\textsf{e}   \ottsym{)}  \ottsym{:}  \ottnt{C}}{%
{\ottdrulename{tye}}{}%
}}

\newcommand{\ottdefntype}[1]{\begin{ottdefnblock}[#1]{$\Gamma  \vdash  \ottnt{e}  \ottsym{:}  \ottnt{C}$}{\ottcom{Expression Typing}}
\ottusedrule{\ottdruletye{}}
\end{ottdefnblock}}

\newcommand{\ottdruletycXXapp}[1]{\ottdrule[#1]{%
\ottpremise{\Gamma  \vdash  \ottnt{v}  \ottsym{:}  \ottnt{A}}%
\ottpremise{\Gamma  \vdash  \ottnt{e}  \ottsym{:}  \ottnt{A}  \rightarrow  \ottnt{C}}%
}{
\Gamma  \vdash  \ottnt{e} \, \ottnt{v}  \ottsym{:}  \ottnt{C}}{%
{\ottdrulename{tyc\_app}}{}%
}}

\newcommand{\ottdruletycXXlam}[1]{\ottdrule[#1]{%
\ottpremise{\Gamma  \ottsym{,}  x  \ottsym{:}  \ottnt{A}  \vdash  \ottnt{e}  \ottsym{:}  \ottnt{C}}%
}{
\Gamma  \vdash   \lambda  x .  \ottnt{e}   \ottsym{:}  \ottnt{A}  \rightarrow  \ottnt{C}}{%
{\ottdrulename{tyc\_lam}}{}%
}}

\newcommand{\ottdruletycXXlet}[1]{\ottdrule[#1]{%
\ottpremise{\Gamma  \vdash  \ottnt{e}  \ottsym{:}  \ottkw{F} \, \ottnt{A}}%
\ottpremise{\Gamma  \ottsym{,}  x  \ottsym{:}  \ottnt{A}  \vdash  \ottnt{e}  \ottsym{:}  \ottnt{C}}%
}{
\Gamma  \vdash   \ottkw{let} \; x \,{\texttt{=} }\, \ottnt{e_{{\mathrm{1}}}} \, \ottkw{in} \, \ottnt{e_{{\mathrm{2}}}}   \ottsym{:}  \ottnt{C}}{%
{\ottdrulename{tyc\_let}}{}%
}}

\newcommand{\ottdruletycXXret}[1]{\ottdrule[#1]{%
\ottpremise{\Gamma  \vdash  \ottnt{v}  \ottsym{:}  \ottnt{A}}%
}{
\Gamma  \vdash  \ottkw{ret} \, \ottnt{v}  \ottsym{:}  \ottkw{F} \, \ottnt{A}}{%
{\ottdrulename{tyc\_ret}}{}%
}}

\newcommand{\ottdruletycXXforce}[1]{\ottdrule[#1]{%
\ottpremise{\Gamma  \vdash  \ottnt{v}  \ottsym{:}  \ottkw{U} \, \ottnt{C}}%
}{
\Gamma  \vdash  \ottkw{force} \, \ottnt{v}  \ottsym{:}  \ottnt{C}}{%
{\ottdrulename{tyc\_force}}{}%
}}

\newcommand{\ottdruletycXXref}[1]{\ottdrule[#1]{%
\ottpremise{\Gamma  \vdash  \ottnt{v}  \ottsym{:}  \ottnt{A}}%
}{
\Gamma  \vdash  \ottkw{ref} \, \ottnt{v}  \ottsym{:}  \ottkw{F} \, \ottsym{(}  \ottkw{Ref} \, \ottnt{A}  \ottsym{)}}{%
{\ottdrulename{tyc\_ref}}{}%
}}

\newcommand{\ottdruletycXXget}[1]{\ottdrule[#1]{%
\ottpremise{\Gamma  \vdash  \ottnt{v}  \ottsym{:}  \ottkw{Ref} \, \ottnt{A}}%
}{
\Gamma  \vdash  \ottkw{get} \, \ottnt{v}  \ottsym{:}  \ottkw{F} \, \ottnt{A}}{%
{\ottdrulename{tyc\_get}}{}%
}}

\newcommand{\ottdruletycXXset}[1]{\ottdrule[#1]{%
\ottpremise{\Gamma  \vdash  \ottnt{v_{{\mathrm{1}}}}  \ottsym{:}  \ottkw{Ref} \, \ottnt{A}}%
\ottpremise{\Gamma  \vdash  \ottnt{v_{{\mathrm{2}}}}  \ottsym{:}  \ottnt{A}}%
}{
\Gamma  \vdash  \ottkw{set} \, \ottnt{v_{{\mathrm{1}}}} \, \ottnt{v_{{\mathrm{2}}}}  \ottsym{:}  \ottkw{F} \, \ottsym{1}}{%
{\ottdrulename{tyc\_set}}{}%
}}

\newcommand{\ottdruletycXXext}[1]{\ottdrule[#1]{%
\ottpremise{\Gamma  \vdash  \ottnt{v_{{\mathrm{1}}}}  \ottsym{:}  \ottkw{Dict} \, \Delta}%
\ottpremise{\Gamma  \vdash  \ottnt{v_{{\mathrm{2}}}}  \ottsym{:}  \ottnt{A}}%
\ottpremise{\Gamma  \vdash  \ottnt{v_{{\mathrm{3}}}}  \ottsym{:}  \ottnt{B}}%
}{
\Gamma  \vdash  \ottkw{ext} \, \ottnt{v_{{\mathrm{1}}}} \, \ottnt{v_{{\mathrm{2}}}} \, \ottnt{v_{{\mathrm{3}}}}  \ottsym{:}  \ottkw{F} \, \ottsym{(}  \ottkw{Dict} \, \ottsym{(}   \Delta ,  \ottnt{v_{{\mathrm{2}}}} \mapsto \ottnt{B}   \ottsym{)}  \ottsym{)}}{%
{\ottdrulename{tyc\_ext}}{}%
}}

\newcommand{\ottdruletycXXprojTop}[1]{\ottdrule[#1]{%
\ottpremise{\Gamma  \vdash  \ottnt{v_{{\mathrm{1}}}}  \ottsym{:}  \textsf{?}}%
\ottpremise{\Gamma  \vdash  \ottnt{v_{{\mathrm{2}}}}  \ottsym{:}  \ottnt{A}}%
}{
\Gamma  \vdash   \ottnt{v_{{\mathrm{1}}}} \left[  \ottnt{v_{{\mathrm{2}}}}  \right]_{  \texttt{?}  }   \ottsym{:}  \ottkw{F} \, \textsf{?}}{%
{\ottdrulename{tyc\_projTop}}{}%
}}

\newcommand{\ottdruletycXXproj}[1]{\ottdrule[#1]{%
\ottpremise{\Gamma  \vdash  \ottnt{v_{{\mathrm{1}}}}  \ottsym{:}  \ottkw{Dict} \, \ottsym{(}   \Delta ,  \ottnt{v_{{\mathrm{2}}}} \mapsto \ottnt{B}   \ottsym{)}}%
\ottpremise{\varepsilon  \vdash  \ottnt{v_{{\mathrm{2}}}}  \ottsym{:}  \ottnt{A}}%
}{
\Gamma  \vdash   \ottnt{v_{{\mathrm{1}}}} \left[  \ottnt{v_{{\mathrm{2}}}}  \right]_{  \texttt{!}  }   \ottsym{:}  \ottkw{F} \, \ottnt{B}}{%
{\ottdrulename{tyc\_proj}}{}%
}}

\newcommand{\ottdruletycXXopenDb}[1]{\ottdrule[#1]{%
\ottpremise{\Gamma  \vdash  \ottnt{v}  \ottsym{:}  \ottkw{Str}}%
}{
\Gamma  \vdash   \texttt{openDb}_{  \texttt{?}  }~ \ottnt{v}   \ottsym{:}  \ottkw{F} \, \ottsym{(}  \ottkw{Db} \, \textsf{?}  \ottsym{)}}{%
{\ottdrulename{tyc\_openDb}}{}%
}}

\newcommand{\ottdruletycXXfilterDb}[1]{\ottdrule[#1]{%
\ottpremise{\Gamma  \vdash  \ottnt{v_{{\mathrm{1}}}}  \ottsym{:}  \ottkw{Db} \, \ottnt{A}}%
\ottpremise{\Gamma  \vdash  \ottnt{v_{{\mathrm{2}}}}  \ottsym{:}  \ottkw{U} \, \ottsym{(}  \ottnt{A}  \rightarrow  \ottkw{F} \, \ottsym{2}  \ottsym{)}}%
}{
\Gamma  \vdash   \texttt{filterDb}_{  \texttt{!}  }~ \ottnt{v_{{\mathrm{1}}}} ~ \ottnt{v_{{\mathrm{2}}}}   \ottsym{:}  \ottkw{F} \, \ottsym{(}  \ottkw{Db} \, \ottnt{A}  \ottsym{)}}{%
{\ottdrulename{tyc\_filterDb}}{}%
}}

\newcommand{\ottdruletycXXjoinDbTop}[1]{\ottdrule[#1]{%
\ottpremise{\Gamma  \vdash  \ottnt{v_{{\mathrm{1}}}}  \ottsym{:}  \ottkw{Db} \, \textsf{?}}%
\ottpremise{\Gamma  \vdash  \ottnt{v_{{\mathrm{2}}}}  \ottsym{:}  \ottnt{A_{{\mathrm{1}}}}}%
\ottpremise{\Gamma  \vdash  \ottnt{v_{{\mathrm{3}}}}  \ottsym{:}  \ottkw{Db} \, \textsf{?}}%
\ottpremise{\Gamma  \vdash  \ottnt{v_{{\mathrm{4}}}}  \ottsym{:}  \ottnt{A_{{\mathrm{2}}}}}%
}{
\Gamma  \vdash   \texttt{joinDb}_{  \texttt{?}  }~ \ottnt{v_{{\mathrm{1}}}} ~ \ottnt{v_{{\mathrm{2}}}} ~ \ottnt{v_{{\mathrm{3}}}} ~ \ottnt{v_{{\mathrm{4}}}}   \ottsym{:}  \ottkw{F} \, \ottsym{(}  \ottkw{Db} \, \textsf{?}  \ottsym{)}}{%
{\ottdrulename{tyc\_joinDbTop}}{}%
}}

\newcommand{\ottdruletycXXjoinDb}[1]{\ottdrule[#1]{%
\ottpremise{\Gamma  \vdash  \ottnt{v_{{\mathrm{1}}}}  \ottsym{:}  \ottkw{Db} \, \ottsym{(}  \ottkw{Dict} \, \ottsym{(}   \Delta_{{\mathrm{1}}} ,  \ottnt{v_{{\mathrm{2}}}} \mapsto \ottnt{B}   \ottsym{)}  \ottsym{)}}%
\ottpremise{\varepsilon  \vdash  \ottnt{v_{{\mathrm{2}}}}  \ottsym{:}  \ottnt{A_{{\mathrm{1}}}}}%
\ottpremise{\Gamma  \vdash  \ottnt{v_{{\mathrm{3}}}}  \ottsym{:}  \ottkw{Db} \, \ottsym{(}  \ottkw{Dict} \, \ottsym{(}   \Delta_{{\mathrm{2}}} ,  \ottnt{v_{{\mathrm{4}}}} \mapsto \ottnt{B}   \ottsym{)}  \ottsym{)}}%
\ottpremise{\varepsilon  \vdash  \ottnt{v_{{\mathrm{4}}}}  \ottsym{:}  \ottnt{A_{{\mathrm{2}}}}}%
}{
\Gamma  \vdash   \texttt{joinDb}_{  \texttt{!}  }~ \ottnt{v_{{\mathrm{1}}}} ~ \ottnt{v_{{\mathrm{2}}}} ~ \ottnt{v_{{\mathrm{3}}}} ~ \ottnt{v_{{\mathrm{4}}}}   \ottsym{:}  \ottkw{F} \, \ottsym{(}  \ottkw{Db} \, \ottsym{(}  \ottkw{Dict} \, \ottsym{(}   \Delta_{{\mathrm{1}}}  ,    \Delta_{{\mathrm{2}}} ,  \ottnt{v_{{\mathrm{2}}}} \mapsto \ottnt{B}  ,  \ottnt{v_{{\mathrm{4}}}} \mapsto \ottnt{B}    \ottsym{)}  \ottsym{)}  \ottsym{)}}{%
{\ottdrulename{tyc\_joinDb}}{}%
}}

\newcommand{\ottdruletycXXAnnot}[1]{\ottdrule[#1]{%
\ottpremise{\Gamma  \vdash  \ottnt{e}  \ottsym{:}  \ottnt{C}}%
}{
\Gamma  \vdash  \ottnt{e}  \mathrel{ \texttt{?:} }  \ottnt{C}  \ottsym{:}  \ottnt{C}}{%
{\ottdrulename{tyc\_Annot}}{}%
}}

\newcommand{\ottdefntypc}[1]{\begin{ottdefnblock}[#1]{$\Gamma  \vdash  \dot{e}  \ottsym{:}  \ottnt{C}$}{\ottcom{Computation Typing}}
\ottusedrule{\ottdruletycXXapp{}}
\ottusedrule{\ottdruletycXXlam{}}
\ottusedrule{\ottdruletycXXlet{}}
\ottusedrule{\ottdruletycXXret{}}
\ottusedrule{\ottdruletycXXforce{}}
\ottusedrule{\ottdruletycXXref{}}
\ottusedrule{\ottdruletycXXget{}}
\ottusedrule{\ottdruletycXXset{}}
\ottusedrule{\ottdruletycXXext{}}
\ottusedrule{\ottdruletycXXprojTop{}}
\ottusedrule{\ottdruletycXXproj{}}
\ottusedrule{\ottdruletycXXopenDb{}}
\ottusedrule{\ottdruletycXXfilterDb{}}
\ottusedrule{\ottdruletycXXjoinDbTop{}}
\ottusedrule{\ottdruletycXXjoinDb{}}
\ottusedrule{\ottdruletycXXAnnot{}}
\end{ottdefnblock}}

\newcommand{\ottdruleordXXret}[1]{\ottdrule[#1]{%
\ottpremise{ \ottnt{A}  \approx  \ottnt{B} }%
}{
 \ottkw{F} \, \ottnt{A}  \approx  \ottkw{F} \, \ottnt{B} }{%
{\ottdrulename{ord\_ret}}{}%
}}

\newcommand{\ottdruleordXXarr}[1]{\ottdrule[#1]{%
\ottpremise{ \ottnt{B}  \approx  \ottnt{A} }%
\ottpremise{ \ottnt{C}  \approx  \ottnt{D} }%
}{
 \ottnt{A}  \rightarrow  \ottnt{C}  \approx  \ottnt{B}  \rightarrow  \ottnt{D} }{%
{\ottdrulename{ord\_arr}}{}%
}}

\newcommand{\ottdefnconsis}[1]{\begin{ottdefnblock}[#1]{$ \ottnt{C}  \approx  \ottnt{D} $}{}
\ottusedrule{\ottdruleordXXret{}}
\ottusedrule{\ottdruleordXXarr{}}
\end{ottdefnblock}}

\newcommand{\ottdruleordXXTopR}[1]{\ottdrule[#1]{%
}{
 \ottnt{A}  \approx  \textsf{?} }{%
{\ottdrulename{ord\_TopR}}{}%
}}

\newcommand{\ottdruleordXXTopL}[1]{\ottdrule[#1]{%
}{
 \textsf{?}  \approx  \ottnt{A} }{%
{\ottdrulename{ord\_TopL}}{}%
}}

\newcommand{\ottdruleordXXDb}[1]{\ottdrule[#1]{%
\ottpremise{ \ottnt{A}  \approx  \ottnt{B} }%
}{
 \ottkw{Db} \, \ottnt{A}  \approx  \ottkw{Db} \, \ottnt{B} }{%
{\ottdrulename{ord\_Db}}{}%
}}

\newcommand{\ottdruleordXXDictEmp}[1]{\ottdrule[#1]{%
}{
 \ottkw{Dict} \, \ottsym{(}  \varepsilon  \ottsym{)}  \approx  \ottkw{Dict} \, \ottsym{(}  \Delta  \ottsym{)} }{%
{\ottdrulename{ord\_DictEmp}}{}%
}}

\newcommand{\ottdruleordXXDictCons}[1]{\ottdrule[#1]{%
\ottpremise{ \ottnt{A}  \approx  \ottnt{B} }%
\ottpremise{ \ottkw{Dict} \, \ottsym{(}  \Delta  \ottsym{)}  \approx  \ottkw{Dict} \, \ottsym{(}  \Delta'  \ottsym{)} }%
}{
 \ottkw{Dict} \, \ottsym{(}   \Delta ,  \ottnt{v_{{\mathrm{1}}}} \mapsto \ottnt{A}   \ottsym{)}  \approx  \ottkw{Dict} \, \ottsym{(}   \Delta' ,  \ottnt{v_{{\mathrm{1}}}} \mapsto \ottnt{B}   \ottsym{)} }{%
{\ottdrulename{ord\_DictCons}}{}%
}}

\newcommand{\ottdruleordXXthunk}[1]{\ottdrule[#1]{%
\ottpremise{ \ottnt{C}  \approx  \ottnt{D} }%
}{
 \ottkw{U} \, \ottnt{C}  \approx  \ottkw{U} \, \ottnt{D} }{%
{\ottdrulename{ord\_thunk}}{}%
}}

\newcommand{\ottdefntopordv}[1]{\begin{ottdefnblock}[#1]{$ \ottnt{A}  \approx  \ottnt{B} $}{}
\ottusedrule{\ottdruleordXXTopR{}}
\ottusedrule{\ottdruleordXXTopL{}}
\ottusedrule{\ottdruleordXXDb{}}
\ottusedrule{\ottdruleordXXDictEmp{}}
\ottusedrule{\ottdruleordXXDictCons{}}
\ottusedrule{\ottdruleordXXthunk{}}
\end{ottdefnblock}}


\newcommand{\ottdefnchkv}[1]{\begin{ottdefnblock}[#1]{$ \Gamma   \vdash   \ottnt{v}  \Leftarrow  \ottnt{A} $}{\ottcom{Value Checking}}
\end{ottdefnblock}}

\newcommand{\ottdruletyv}[1]{\ottdrule[#1]{%
\ottpremise{\Gamma  \vdash  \dot{v}  \ottsym{:}  \ottnt{A}}%
}{
\Gamma  \vdash  \ottsym{(}   \dot{v} ~\texttt{@:}~ a^\textsf{v}   \ottsym{)}  \ottsym{:}  \ottnt{A}}{%
{\ottdrulename{tyv}}{}%
}}

\newcommand{\ottdefntyv}[1]{\begin{ottdefnblock}[#1]{$\Gamma  \vdash  \ottnt{v}  \ottsym{:}  \ottnt{A}$}{\ottcom{Value Typing}}
\ottusedrule{\ottdruletyv{}}
\end{ottdefnblock}}

\newcommand{\ottdruletypvXXvar}[1]{\ottdrule[#1]{%
}{
\Gamma  \ottsym{,}  x  \ottsym{:}  \ottnt{A}  \vdash  x  \ottsym{:}  \ottnt{A}}{%
{\ottdrulename{typv\_var}}{}%
}}

\newcommand{\ottdruletypvXXloc}[1]{\ottdrule[#1]{%
}{
 \left|  \mu  ,  \ell \mapsto \ottsym{(}   \dot{v} ~\texttt{@:}~ \ottnt{A}   \ottsym{)}  \right|   \ottsym{,}  \Gamma  \vdash  \ottkw{loc} \, \ell  \ottsym{:}  \ottnt{A}}{%
{\ottdrulename{typv\_loc}}{}%
}}

\newcommand{\ottdruletypvXXStr}[1]{\ottdrule[#1]{%
}{
\Gamma  \vdash  \ottkw{str} \, \ottmv{s}  \ottsym{:}  \ottkw{Str}}{%
{\ottdrulename{typv\_Str}}{}%
}}

\newcommand{\ottdruletypvXXNum}[1]{\ottdrule[#1]{%
}{
\Gamma  \vdash  \ottkw{num} \, \ottmv{n}  \ottsym{:}  \ottkw{Num}}{%
{\ottdrulename{typv\_Num}}{}%
}}

\newcommand{\ottdruletypvXXdictEmp}[1]{\ottdrule[#1]{%
}{
\Gamma  \vdash  \ottkw{dict} \, \varepsilon  \ottsym{:}  \ottkw{Dict} \, \varepsilon}{%
{\ottdrulename{typv\_dictEmp}}{}%
}}

\newcommand{\ottdruletypvXXdictCons}[1]{\ottdrule[#1]{%
\ottpremise{\Gamma  \vdash  \ottkw{dict} \, \ottsym{(}  \delta  \ottsym{)}  \ottsym{:}  \ottkw{Dict} \, \ottsym{(}  \Delta  \ottsym{)}}%
\ottpremise{\Gamma  \vdash  \ottnt{v_{{\mathrm{1}}}}  \ottsym{:}  \ottnt{A}}%
\ottpremise{\Gamma  \vdash  \ottnt{v_{{\mathrm{2}}}}  \ottsym{:}  \ottnt{B}}%
}{
\Gamma  \vdash  \ottkw{dict} \, \ottsym{(}   \delta  ,  \ottnt{v_{{\mathrm{1}}}}  \mapsto  \ottnt{v_{{\mathrm{2}}}}   \ottsym{)}  \ottsym{:}  \ottkw{Dict} \, \ottsym{(}   \Delta ,  \ottnt{v_{{\mathrm{1}}}} \mapsto \ottnt{B}   \ottsym{)}}{%
{\ottdrulename{typv\_dictCons}}{}%
}}

\newcommand{\ottdruletypvXXothunk}[1]{\ottdrule[#1]{%
\ottpremise{\Gamma  \vdash  \ottnt{e}  \ottsym{:}  \ottnt{C}}%
}{
\Gamma  \vdash  \ottkw{othunk} \, \ottnt{e}  \ottsym{:}  \ottkw{U} \, \ottnt{C}}{%
{\ottdrulename{typv\_othunk}}{}%
}}

\newcommand{\ottdruletypvXXthunk}[1]{\ottdrule[#1]{%
\ottpremise{\Gamma  \vdash  \rho  \leadsto  \Gamma'}%
\ottpremise{\Gamma'  \vdash  \ottnt{e}  \ottsym{:}  \ottnt{C}}%
}{
\Gamma  \vdash  \ottkw{thunk} \, \rho \, \ottnt{e}  \ottsym{:}  \ottkw{U} \, \ottnt{C}}{%
{\ottdrulename{typv\_thunk}}{}%
}}

\newcommand{\ottdefntypv}[1]{\begin{ottdefnblock}[#1]{$\Gamma  \vdash  \dot{v}  \ottsym{:}  \ottnt{A}$}{\ottcom{Value Typing}}
\ottusedrule{\ottdruletypvXXvar{}}
\ottusedrule{\ottdruletypvXXloc{}}
\ottusedrule{\ottdruletypvXXStr{}}
\ottusedrule{\ottdruletypvXXNum{}}
\ottusedrule{\ottdruletypvXXdictEmp{}}
\ottusedrule{\ottdruletypvXXdictCons{}}
\ottusedrule{\ottdruletypvXXothunk{}}
\ottusedrule{\ottdruletypvXXthunk{}}
\end{ottdefnblock}}

\newcommand{\ottdruleenvtXXemp}[1]{\ottdrule[#1]{%
}{
\Gamma  \vdash  \varepsilon  \leadsto  \varepsilon}{%
{\ottdrulename{envt\_emp}}{}%
}}

\newcommand{\ottdruleenvtXXcons}[1]{\ottdrule[#1]{%
\ottpremise{\Gamma  \vdash  \rho  \leadsto  \Gamma'}%
\ottpremise{\Gamma  \vdash  \ottnt{v}  \ottsym{:}  \ottnt{A}}%
}{
\Gamma  \vdash   \rho  ,  x \mapsto \ottnt{v}   \leadsto  \Gamma'  \ottsym{,}  x  \ottsym{:}  \ottnt{A}}{%
{\ottdrulename{envt\_cons}}{}%
}}

\newcommand{\ottdefnenvt}[1]{\begin{ottdefnblock}[#1]{$\Gamma  \vdash  \rho  \leadsto  \Gamma'$}{\ottcom{Environment Typing}}
\ottusedrule{\ottdruleenvtXXemp{}}
\ottusedrule{\ottdruleenvtXXcons{}}
\end{ottdefnblock}}

\newcommand{\ottdrulestacktXXemp}[1]{\ottdrule[#1]{%
}{
 \Gamma   \vdash   \textsf{halt}  \Leftarrow  \ottnt{C} }{%
{\ottdrulename{stackt\_emp}}{}%
}}

\newcommand{\ottdrulestacktXXlet}[1]{\ottdrule[#1]{%
\ottpremise{ \Gamma_{{\mathrm{1}}}   \vdash   \kappa  \Leftarrow  \ottnt{C} }%
\ottpremise{\Gamma_{{\mathrm{1}}}  \vdash  \rho  \leadsto  \Gamma_{{\mathrm{2}}}}%
\ottpremise{\Gamma_{{\mathrm{2}}}  \ottsym{,}  x  \ottsym{:}  \ottnt{A}  \vdash  \ottnt{e}  \ottsym{:}  \ottnt{C}}%
}{
 \Gamma_{{\mathrm{1}}}   \vdash   \kappa  ::  \ottsym{(}  \rho  \ottsym{,}  x  \ottsym{.}  \ottnt{e}  \ottsym{)}  \Leftarrow  \ottkw{F} \, \ottnt{A} }{%
{\ottdrulename{stackt\_let}}{}%
}}

\newcommand{\ottdrulestacktXXapp}[1]{\ottdrule[#1]{%
\ottpremise{\Gamma  \vdash  \ottnt{v}  \ottsym{:}  \ottnt{A}}%
\ottpremise{ \Gamma   \vdash   \kappa  \Leftarrow  \ottnt{C} }%
}{
 \Gamma   \vdash   \kappa  ::  \ottnt{v}  \Leftarrow  \ottnt{A}  \rightarrow  \ottnt{C} }{%
{\ottdrulename{stackt\_app}}{}%
}}

\newcommand{\ottdefnstackt}[1]{\begin{ottdefnblock}[#1]{$ \Gamma   \vdash   \kappa  \Leftarrow  \ottnt{C} $}{\ottcom{Stack Typing}}
\ottusedrule{\ottdrulestacktXXemp{}}
\ottusedrule{\ottdrulestacktXXlet{}}
\ottusedrule{\ottdrulestacktXXapp{}}
\end{ottdefnblock}}

\newcommand{\ottdruleStateXXok}[1]{\ottdrule[#1]{%
\ottpremise{ \left| \mu \right|   \vdash  \rho  \leadsto  \Gamma}%
\ottpremise{ \left| \mu \right|   \ottsym{,}  \Gamma  \vdash  \dot{e}  \ottsym{:}  \ottnt{C}}%
\ottpremise{  \left| \mu \right|    \vdash   \kappa  \Leftarrow  \ottnt{C} }%
}{
 \left<  \mu ;  \kappa ;  \rho ;  \dot{e}  \right>  \, {\textsf{ok} }}{%
{\ottdrulename{State\_ok}}{}%
}}

\newcommand{\ottdefnstateXXok}[1]{\begin{ottdefnblock}[#1]{$\sigma \, {\textsf{ok} }$}{}
\ottusedrule{\ottdruleStateXXok{}}
\end{ottdefnblock}}

\newcommand{\ottdruleFinXXret}[1]{\ottdrule[#1]{%
}{
 \left<  \mu ;  \textsf{halt} ;  \rho ;  \ottkw{ret} \, \ottnt{v}  \right>  \, \ottkw{final}}{%
{\ottdrulename{Fin\_ret}}{}%
}}

\newcommand{\ottdruleFinXXlam}[1]{\ottdrule[#1]{%
}{
 \left<  \mu ;  \textsf{halt} ;  \rho ;   \lambda  x .  \ottnt{e}   \right>  \, \ottkw{final}}{%
{\ottdrulename{Fin\_lam}}{}%
}}

\newcommand{\ottdefnstateXXfinal}[1]{\begin{ottdefnblock}[#1]{$\sigma \, \ottkw{final}$}{}
\ottusedrule{\ottdruleFinXXret{}}
\ottusedrule{\ottdruleFinXXlam{}}
\end{ottdefnblock}}


\newcommand{\ottdefnfindXXvar}[1]{\begin{ottdefnblock}[#1]{$ \rho ( x ) =  \ottnt{v} $}{}
\end{ottdefnblock}}


\newcommand{\ottdefncloseXXdict}[1]{\begin{ottdefnblock}[#1]{$ \rho ( \delta ) \leadsto  \delta' $}{}
\end{ottdefnblock}}

\newcommand{\ottdruleSXXvar}[1]{\ottdrule[#1]{%
\ottpremise{ \rho ( x ) =  \ottnt{v} }%
}{
 \rho ( \ottsym{(}   x ~\texttt{@:}~ a^\textsf{v}   \ottsym{)} ) \leadsto  \ottnt{v} }{%
{\ottdrulename{S\_var}}{}%
}}

\newcommand{\ottdruleSXXdict}[1]{\ottdrule[#1]{%
\ottpremise{ \rho ( \delta ) \leadsto  \delta' }%
}{
 \rho ( \ottsym{(}   \ottkw{dict} \, \delta ~\texttt{@:}~ a^\textsf{v}   \ottsym{)} ) \leadsto  \ottsym{(}   \ottkw{dict} \, \delta' ~\texttt{@:}~ a^\textsf{v}   \ottsym{)} }{%
{\ottdrulename{S\_dict}}{}%
}}

\newcommand{\ottdruleSXXnum}[1]{\ottdrule[#1]{%
}{
 \rho ( \ottsym{(}   \ottkw{num} \, \ottmv{n} ~\texttt{@:}~ a^\textsf{v}   \ottsym{)} ) \leadsto  \ottsym{(}   \ottkw{num} \, \ottmv{n} ~\texttt{@:}~ a^\textsf{v}   \ottsym{)} }{%
{\ottdrulename{S\_num}}{}%
}}

\newcommand{\ottdruleSXXstr}[1]{\ottdrule[#1]{%
}{
 \rho ( \ottsym{(}   \ottkw{str} \, \ottmv{s} ~\texttt{@:}~ a^\textsf{v}   \ottsym{)} ) \leadsto  \ottsym{(}   \ottkw{str} \, \ottmv{s} ~\texttt{@:}~ a^\textsf{v}   \ottsym{)} }{%
{\ottdrulename{S\_str}}{}%
}}

\newcommand{\ottdruleSXXloc}[1]{\ottdrule[#1]{%
}{
 \rho ( \ottsym{(}   \ottkw{loc} \, \ell ~\texttt{@:}~ a^\textsf{v}   \ottsym{)} ) \leadsto  \ottsym{(}   \ottkw{loc} \, \ell ~\texttt{@:}~ a^\textsf{v}   \ottsym{)} }{%
{\ottdrulename{S\_loc}}{}%
}}

\newcommand{\ottdruleSXXothunk}[1]{\ottdrule[#1]{%
}{
 \rho ( \ottsym{(}   \ottkw{othunk} \, \ottnt{e} ~\texttt{@:}~ a^\textsf{v}   \ottsym{)} ) \leadsto  \ottsym{(}   \ottkw{thunk} \, \rho \, \ottnt{e} ~\texttt{@:}~ a^\textsf{v}   \ottsym{)} }{%
{\ottdrulename{S\_othunk}}{}%
}}

\newcommand{\ottdruleSXXthunk}[1]{\ottdrule[#1]{%
}{
 \rho ( \ottsym{(}   \ottkw{thunk} \, \rho' \, \ottnt{e} ~\texttt{@:}~ a^\textsf{v}   \ottsym{)} ) \leadsto  \ottsym{(}   \ottkw{thunk} \, \rho' \, \ottnt{e} ~\texttt{@:}~ a^\textsf{v}   \ottsym{)} }{%
{\ottdrulename{S\_thunk}}{}%
}}

\newcommand{\ottdefncloseXXpv}[1]{\begin{ottdefnblock}[#1]{$ \rho ( \ottnt{v} ) \leadsto  \ottnt{v'} $}{}
\ottusedrule{\ottdruleSXXvar{}}
\ottusedrule{\ottdruleSXXdict{}}
\ottusedrule{\ottdruleSXXnum{}}
\ottusedrule{\ottdruleSXXstr{}}
\ottusedrule{\ottdruleSXXloc{}}
\ottusedrule{\ottdruleSXXothunk{}}
\ottusedrule{\ottdruleSXXthunk{}}
\end{ottdefnblock}}


\newcommand{\ottdefnlocXXnotXXinXXstore}[1]{\begin{ottdefnblock}[#1]{$ \ell  \not\in  \mu $}{}
\end{ottdefnblock}}


\newcommand{\ottdefndometaXXstep}[1]{\begin{ottdefnblock}[#1]{$ [\![  \sigma  ]\!] \vdash  e_\textsf{m}  \Downarrow_{\textsf{rcc} } state' $}{}
\end{ottdefnblock}}

\newcommand{\ottdruleSXXchk}[1]{\ottdrule[#1]{%
\ottpremise{ [\![   \left<  \mu ;  \kappa ;  \rho ;  \dot{e}  \right>   ]\!] \vdash  e_\textsf{m}  \Downarrow_{\textsf{rcc} } state' }%
}{
 \left<  \mu ;  \kappa ;  \rho ;  \ottkw{rcc} \, e_\textsf{m} \, \ottsym{(}   \dot{e} ~\texttt{@:}~ a^\textsf{e}   \ottsym{)}  \right>   \longrightarrow  \sigma'}{%
{\ottdrulename{S\_chk}}{}%
}}

\newcommand{\ottdruleSXXlet}[1]{\ottdrule[#1]{%
}{
 \left<  \mu ;  \kappa ;  \rho ;   \ottkw{let} \; x \,{\texttt{=} }\, \ottsym{(}   \dot{e}_{{\mathrm{1}}} ~\texttt{@:}~ a^\textsf{e}_{{\mathrm{1}}}   \ottsym{)} \, \ottkw{in} \, \ottnt{e_{{\mathrm{2}}}}   \right>   \longrightarrow   \left<  \mu ;  \kappa  ::  \ottsym{(}  \rho  \ottsym{,}  x  \ottsym{.}  \ottnt{e_{{\mathrm{2}}}}  \ottsym{)} ;  \rho ;  \dot{e}_{{\mathrm{1}}}  \right> }{%
{\ottdrulename{S\_let}}{}%
}}

\newcommand{\ottdruleSXXret}[1]{\ottdrule[#1]{%
\ottpremise{ \rho ( \ottnt{v} ) \leadsto  \ottnt{v'} }%
}{
 \left<  \mu ;  \kappa  ::  \ottsym{(}  \rho  \ottsym{,}  x  \ottsym{.}  \ottsym{(}   \dot{e} ~\texttt{@:}~ a^\textsf{e}   \ottsym{)}  \ottsym{)} ;  \rho' ;  \ottkw{ret} \, \ottnt{v}  \right>   \longrightarrow   \left<  \mu ;  \kappa ;   \rho  ,  x \mapsto \ottnt{v'}  ;  \dot{e}  \right> }{%
{\ottdrulename{S\_ret}}{}%
}}

\newcommand{\ottdruleSXXapp}[1]{\ottdrule[#1]{%
\ottpremise{ \rho ( \ottnt{v} ) \leadsto  \ottnt{v'} }%
}{
 \left<  \mu ;  \kappa ;  \rho ;  \ottsym{(}   \dot{e} ~\texttt{@:}~ a^\textsf{e}   \ottsym{)} \, \ottnt{v}  \right>   \longrightarrow   \left<  \mu ;  \kappa  ::  \ottnt{v'} ;  \rho ;  \dot{e}  \right> }{%
{\ottdrulename{S\_app}}{}%
}}

\newcommand{\ottdruleSXXlam}[1]{\ottdrule[#1]{%
}{
 \left<  \mu ;  \kappa  ::  \ottnt{v} ;  \rho ;   \lambda  x .  \ottsym{(}   \dot{e} ~\texttt{@:}~ a^\textsf{e}   \ottsym{)}   \right>   \longrightarrow   \left<  \mu ;  \kappa ;   \rho  ,  x \mapsto \ottnt{v}  ;  \dot{e}  \right> }{%
{\ottdrulename{S\_lam}}{}%
}}

\newcommand{\ottdruleSXXforce}[1]{\ottdrule[#1]{%
\ottpremise{ \rho ( \ottnt{v} ) \leadsto  \ottsym{(}   \ottkw{thunk} \, \rho' \, \ottsym{(}   \dot{e} ~\texttt{@:}~ a^\textsf{e}   \ottsym{)} ~\texttt{@:}~ a^\textsf{v}   \ottsym{)} }%
}{
 \left<  \mu ;  \kappa ;  \rho ;  \ottkw{force} \, \ottnt{v}  \right>   \longrightarrow   \left<  \mu ;  \kappa ;  \rho' ;  \dot{e}  \right> }{%
{\ottdrulename{S\_force}}{}%
}}

\newcommand{\ottdruleSXXref}[1]{\ottdrule[#1]{%
\ottpremise{ \ell  \not\in  \mu }%
}{
 \left<  \mu ;  \kappa ;  \rho ;  \ottkw{ref} \, \ottnt{v}  \right>   \longrightarrow   \left<   \mu  ,  \ell \mapsto \ottnt{v}  ;  \kappa ;  \rho ;  \ottkw{ret} \, \ottsym{(}   \ottkw{loc} \, \ell ~\texttt{@:}~ \textsf{?}   \ottsym{)}  \right> }{%
{\ottdrulename{S\_ref}}{}%
}}

\newcommand{\ottdruleSXXset}[1]{\ottdrule[#1]{%
\ottpremise{ \rho ( \ottnt{v_{{\mathrm{1}}}} ) \leadsto  \ottsym{(}   \ottkw{loc} \, \ell ~\texttt{@:}~ a^\textsf{v}   \ottsym{)} }%
\ottpremise{ \rho ( \ottnt{v_{{\mathrm{2}}}} ) \leadsto  \ottnt{v'_{{\mathrm{2}}}} }%
}{
 \left<  \mu ;  \kappa ;  \rho ;  \ottkw{set} \, \ottnt{v_{{\mathrm{1}}}} \, \ottnt{v_{{\mathrm{2}}}}  \right>   \longrightarrow   \left<   \mu  ,  \ell \mapsto \ottnt{v'_{{\mathrm{2}}}}  ;  \kappa ;  \rho ;  \ottkw{ret} \, \ottsym{(}   \ottsym{()} ~\texttt{@:}~ \textsf{?}   \ottsym{)}  \right> }{%
{\ottdrulename{S\_set}}{}%
}}

\newcommand{\ottdruleSXXget}[1]{\ottdrule[#1]{%
\ottpremise{ \rho ( \ottnt{v_{{\mathrm{1}}}} ) \leadsto  \ottsym{(}   \ottkw{loc} \, \ell ~\texttt{@:}~ a^\textsf{v}   \ottsym{)} }%
}{
 \left<   \mu  ,  \ell \mapsto \ottnt{v_{{\mathrm{2}}}}  ;  \kappa ;  \rho ;  \ottkw{get} \, \ottnt{v_{{\mathrm{1}}}}  \right>   \longrightarrow   \left<   \mu  ,  \ell \mapsto \ottnt{v_{{\mathrm{2}}}}  ;  \kappa ;  \rho ;  \ottkw{ret} \, \ottnt{v_{{\mathrm{2}}}}  \right> }{%
{\ottdrulename{S\_get}}{}%
}}

\newcommand{\ottdruleSXXext}[1]{\ottdrule[#1]{%
\ottpremise{ \rho ( \ottnt{v_{{\mathrm{1}}}} ) \leadsto  \ottsym{(}   \ottkw{dict} \, \delta ~\texttt{@:}~ a^\textsf{v}   \ottsym{)} }%
\ottpremise{ \rho ( \ottnt{v_{{\mathrm{2}}}} ) \leadsto  \ottnt{v'_{{\mathrm{2}}}} }%
\ottpremise{ \rho ( \ottnt{v_{{\mathrm{3}}}} ) \leadsto  \ottnt{v'_{{\mathrm{3}}}} }%
}{
 \left<  \mu ;  \kappa ;  \rho ;  \ottkw{ext} \, \ottnt{v_{{\mathrm{1}}}} \, \ottnt{v_{{\mathrm{2}}}} \, \ottnt{v_{{\mathrm{3}}}}  \right>   \longrightarrow   \left<  \mu ;  \kappa ;  \rho ;  \ottkw{ret} \, \ottsym{(}   \ottkw{dict} \, \ottsym{(}   \delta  ,  \ottnt{v'_{{\mathrm{2}}}}  \mapsto  \ottnt{v'_{{\mathrm{3}}}}   \ottsym{)} ~\texttt{@:}~ \textsf{?}   \ottsym{)}  \right> }{%
{\ottdrulename{S\_ext}}{}%
}}

\newcommand{\ottdruleSXXproj}[1]{\ottdrule[#1]{%
\ottpremise{ \rho ( \ottnt{v_{{\mathrm{2}}}} ) \leadsto  \ottnt{v'_{{\mathrm{2}}}} }%
\ottpremise{ \rho ( \ottnt{v_{{\mathrm{1}}}} ) \leadsto  \ottsym{(}   \ottkw{dict} \, \ottsym{(}   \delta  ,  \ottnt{v'_{{\mathrm{2}}}}  \mapsto  \ottnt{v_{{\mathrm{3}}}}   \ottsym{)} ~\texttt{@:}~ a^\textsf{v}   \ottsym{)} }%
}{
 \left<  \mu ;  \kappa ;  \rho ;   \ottnt{v_{{\mathrm{1}}}} \left[  \ottnt{v_{{\mathrm{2}}}}  \right]_{ a^{\textsf{op} } }   \right>   \longrightarrow   \left<  \mu ;  \kappa ;  \rho ;  \ottkw{ret} \, \ottnt{v_{{\mathrm{3}}}}  \right> }{%
{\ottdrulename{S\_proj}}{}%
}}

\newcommand{\ottdefnstateXXstep}[1]{\begin{ottdefnblock}[#1]{$\sigma  \longrightarrow  \sigma'$}{}
\ottusedrule{\ottdruleSXXchk{}}
\ottusedrule{\ottdruleSXXlet{}}
\ottusedrule{\ottdruleSXXret{}}
\ottusedrule{\ottdruleSXXapp{}}
\ottusedrule{\ottdruleSXXlam{}}
\ottusedrule{\ottdruleSXXforce{}}
\ottusedrule{\ottdruleSXXref{}}
\ottusedrule{\ottdruleSXXset{}}
\ottusedrule{\ottdruleSXXget{}}
\ottusedrule{\ottdruleSXXext{}}
\ottusedrule{\ottdruleSXXproj{}}
\end{ottdefnblock}}

\newcommand{\ottdefnsJtyp}{
\ottdefndomXXenvXXcontainsXXfvXXe{}\ottdefnvXXclosed{}\ottdefnpvXXclosed{}\ottdefntype{}\ottdefntypc{}\ottdefnconsis{}\ottdefntopordv{}\ottdefnchkv{}\ottdefntyv{}\ottdefntypv{}\ottdefnenvt{}\ottdefnstackt{}\ottdefnstateXXok{}\ottdefnstateXXfinal{}\ottdefnfindXXvar{}\ottdefncloseXXdict{}\ottdefncloseXXpv{}\ottdefnlocXXnotXXinXXstore{}\ottdefndometaXXstep{}\ottdefnstateXXstep{}}

\newcommand{\ottdefnss}{
\ottdefnsJtyp
}

\newcommand{\ottall}{\ottmetavars\\[0pt]
\ottgrammar\\[5.0mm]
\ottdefnss}

\title{A Vision for Online Verification-Validation}

\authorinfo{Matthew A. Hammer}{University of Colorado Boulder}{matthew.hammer@colorado.edu}
\authorinfo{Bor-Yuh Evan Chang}{University of Colorado Boulder}{evan.chang@colorado.edu}
\authorinfo{David Van Horn}{University of Maryland, College Park}{dvanhorn@cs.umd.edu}

\maketitle

\setlength{\pdfpageheight}{\paperheight}
\setlength{\pdfpagewidth}{\paperwidth}

\conferenceinfo{CONF 'yy}{Month d--d, 20yy, City, ST, Country}
\copyrightyear{20yy}
\copyrightdata{978-1-nnnn-nnnn-n/yy/mm}
\copyrightdoi{nnnnnnn.nnnnnnn}



\newcommand{\lambdaVMF}{$\lambda$-\textsf{VMF}\xspace}
\newcommand{\libDb}{\textsf{libDb}\xspace}

\newcommand{\ForallAnalysis}{$\forall$-analysis\xspace}
\newcommand{\ExistsAnalysis}{$\exists$-analysis\xspace}

\newcommand{\ForallAnalyses}{$\forall$-analyses\xspace}
\newcommand{\ExistsAnalyses}{$\exists$-analyses\xspace}

\newcommand{\bnfas}{\ensuremath{\mathrel{::=}}}
\newcommand{\bnfaltbrk}{\hspace{-1.02ex}\bnfalt}
\newenvironment{jgrammar}{\vspace{1pt} \begin{center} ~\!\!\begin{tabular}[t]{@{}lr@{~~}c@{~~}lll@{}}}{\end{tabular}\end{center}}

%
%
%
%
%
%
%
%
%

\def \TirNameStyle #1{{#1}}
\newcommand{\Infer}[3]{\inferrule*[right={\text{\strut#1}}]{{}#2\mathstrut}{{}#3\mathstrut}}

%
%
%
%
%
%
%
%


\newdimen\zzfontsz
\newcommand{\fontsz}[2]{\zzfontsz=#1%
{\fontsize{\zzfontsz}{1.2\zzfontsz}\selectfont{#2}}}
\newcommand{\mathsz}[2]{\text{\fontsz{#1}{$#2$}}}
\newcommand{\textgraybox}[1]{\boxed{#1}}
\newcommand{\graybox}[1]{\textgraybox{\ensuremath{#1}}}
\newcommand{\judgboxfontsize}[1]{\mathsz{11pt}{#1}}
\newcommand{\judgbox}[2]{%
      {\raggedright \textgraybox{\ensuremath{\judgboxfontsize{#1}}}\!\begin{tabular}[c]{l} #2 \end{tabular} %
        \\[-1.5ex]
}}

\begin{abstract}
Today's programmers face a false choice between creating software that
is extensible and software that is correct. Specifically, dynamic
languages permit software that is richly extensible (via dynamic code
loading, dynamic object extension, and various forms of reflection),
and today's programmers exploit this flexibility to ``bring their own
language features'' to enrich extensible languages (e.g., by using
common JavaScript libraries).  Meanwhile, such library-based language
extensions generally lack enforcement of their abstractions, leading
to programming errors that are complex to avoid and predict.

To offer verification for this extensible world, we propose online
verification-validation (OVV), which consists of language and VM
design that enables a ``phaseless'' approach to program analysis, in
contrast to the standard static-dynamic phase distinction.  Phaseless
analysis freely interposes abstract interpretation with concrete
execution, allowing analyses to use dynamic (concrete) information to
prove universal (abstract) properties about future execution.

In this paper, we present a conceptual overview of OVV through a
motivating example program that uses a hypothetical database library.
We present a generic semantics for OVV, and an extension to this
semantics that offers a simple gradual type system for the database
library primitives. The result of instantiating this gradual type
system in an OVV setting is a checker that can progressively type
successive continuations of the program until a continuation is
fully verified. To evaluate the proposed vision of OVV for this
example, we implement the VM semantics (in Rust), and show that this
design permits progressive typing in this manner.
\end{abstract}

\section{Introduction}
\label{sec:introduction}

\JEDI{Problem: Typing the unknown.}  We consider the problem of typing
unknown, dynamically-determined data obtained from the environment. To
illustrate what we mean by unknown, dynamically-determined data,
consider the code in \figref{openDb-authors} that loads a
comma-separated value (CSV) file \lstinline!authors.csv! using the
\code{openDb("authors.csv")} call on
\reftxt{line}{line-openDb-authors}. The contents of this file are
organized into lines, where each line is a row of field values.  The
first line of the file is special and contains a list of field names
instead of field value---see \figref{authors.csv} for example content.
In \reftxt{line}{line-openDb-authors} of
\figref{openDb-authors}, the programmer filters the author
list down to those that have US citizenship using the field
projection \code{author.citizenship}.

Suppose the programmer merely wants to know that their code \emph{will not access undefined fields}----that, in this respect, the program is well-typed.
Given that these object fields are defined by the
\emph{dynamic-generation of data structures} via \code{openDb}, the
validity of this field projection \code{author.citizenship} is
generally unknowable until after \reftxt{line}{line-openDb-authors} when the structure of the
\code{authors} table is defined based on the contents of
\lstinline!authors.csv!.
At the same time, this field projection is clearly valid because of
the special first line in specifically this \lstinline!authors.csv!
shown in \figref{authors.csv}.

\newsavebox{\SBoxOpenDbAuthors}
\begin{lrbox}{\SBoxOpenDbAuthors}
\begin{lstlisting}[style=number]
let authors     = openDb("authors.csv")#\label{line-openDb-authors}#
let authorsUS   = filterDb(authors, (author) =>#\label{line-filterDb}\lststopn#
                           author.citizenship === "US")#\lststartn#
\end{lstlisting}
\end{lrbox}

\newsavebox{\SBoxAuthorsDb}
\begin{lrbox}{\SBoxAuthorsDb}\footnotesize
\begin{lstlisting}[style=default]
name, citizenship, institution
Evan Chang, US, University of Colorado
Roly Perera, UK, University of Glasgow
Matthew Hammer, US, University of Colorado
David Van Horn, US, University of Maryland
\end{lstlisting}
\end{lrbox}

\begin{figure}[tb]\centering
\subfloat[Open a database with an implicit schema and filter the content.]{
  \label{fig:openDb-authors}
  \usebox{\SBoxOpenDbAuthors}
}
\par\subfloat[File~\code{authors.csv}, example content.]{
  \label{fig:authors.csv}
  \usebox{\SBoxAuthorsDb}
}
\caption{Programming with dynamically-determined tables and objects in JavaScript}
\end{figure}

There is a tension here between static and dynamic checking. On one
hand, once the \code{authors} table has been determined, the
programmer would like the field projections from the rows to be
statically verified. But on other hand, how would static checking be
feasible if the schema of \code{authors} database is
dynamically-determined---what would be the type of the \code{openDb} function?

\JEDI{Insight: Pausing concrete execution, using abstract
  interpretation on the continuation.}  To resolve this tension, our
key insight is to imagine pausing the concrete execution of
\figref{openDb-authors} right after
\reftxt{line}{line-openDb-authors} and before
\reftxt{line}{line-filterDb}. Then, imagine reflecting the continuation
of the program's execution to apply abstract interpretation or other
static techniques to prove the validity of \emph{all} subsequent field
projections on the rows of the \code{authors} table.

\JEDI{Our approach: phaseless OVV.}
Our vision centers around a new paradigm for verification and
validation that we call \emph{online verification-validation} (OVV).
In contrast to today's phasic analysis techniques, OVV is
\emph{phaseless}: ``static'' analysis is freely interposed with
``dynamic'' execution.
By virtue of this mixed approach, OVV transcends the conventional
phase distinctions of ``static'' and ``dynamic'' analysis.
To avoid confusion in this phaseless setting, we refer to static analysis
techniques as those for \emph{\ForallAnalysis}, since they demonstrate
\emph{universal} properties of future program's states.
Likewise, we refer to the techniques used in dynamic analysis as those
for \emph{\ExistsAnalysis}, since they demonstrate \emph{existential}
properties about the past program's states.

In this paper, we make the following contributions:
\begin{itemize}
\item We define \lambdaVMF, an abstract machine semantics that enables a form of
  phaseless, online verification-validation (\secref{vm-semantics}).
  The key idea is annotating potentially faulting operations (e.g.,
  field projection) with a \emph{certain} (\texttt{!}) or \emph{uncertain}
  (\textsf{?}) flag, indicating whether or not that operation can be verified.
  Executing uncertain operations gets stuck, so it is up to the OVV program to
  progressively rewrite uncertain operations into certain ones by proving
  the safety of the potentially faulting operation.
\item We present a case study of instantiating online
  ver\-i\-fi\-ca\-tion-validation with a simple, bidirectional gradual type
  system for dynamic field projection and databases with dynamic schemas (\secref{typing}).  The result of
  instantiating a gradual type system in an online
  verification-validation setting is a checker that can
  progressively type successive continuations of the program until
  a continuation is fully verified (i.e., ``statically'' typed).
\item We implement the proposed design for \lambdaVMF in Rust to demonstrate that
  the proposed instantiation of OVV indeed realizes progressive typing.
  Our implementation is public: \url{https://github.com/cuplv/vmfuture}.
\end{itemize}

In the next section, we dig deeper into what we term online
verification-validation by following the \emph{progressive typing} of
the example program from \figref{openDb-authors}.

\section{Overview}
\label{sec:overview}

In \figref{full-example-code}, consider an extension of the example from
\figref{openDb-authors} with two additional lines. The file
\lstinline!books.csv! is opened on \reftxt{line}{line-openDb-books} and contains
a CSV file with book information (shown in \figref{books.csv}). The final line
(\reftxt{line}{line-joinDb}) creates a table of books written by US authors,
along with the information about those authors.

\newsavebox{\SBoxOpenDbAll}
\begin{lrbox}{\SBoxOpenDbAll}
\begin{lstlisting}[style=number]
let authors     = openDb("authors.csv")#\label{line-openDb-authors-full}#
let authorsUS   = filterDb(authors, (author) =>#\label{line-filterDb-full}\lststopn#
                           author.citizenship === "US")#\lststartn#
let books       = openDb("books.csv")#\label{line-openDb-books}#
let authbooksUS = joinDb(authorsUS, "name",#\label{line-joinDb}\lststopn#
                         books, "author")#\lststartn#
\end{lstlisting}
\end{lrbox}
\newsavebox{\SBoxBooksDb}
\begin{lrbox}{\SBoxBooksDb}\footnotesize
\begin{lstlisting}[style=default]
author, title, year, publisher
$\ldots$
\end{lstlisting}
\end{lrbox}
\begin{figure}[tb]\centering
\subfloat[The \code{joinDb(db1,key1,db2,key2)} library function accesses the key
fields using run-time reflection (e.g., \code{db1[i][key1]}).]{
\label{fig:full-example-code}
\usebox{\SBoxOpenDbAll}
}
\par\subfloat[File~\code{books.csv}, example content.]{
\label{fig:books.csv}
\usebox{\SBoxBooksDb}
}
\caption{Continuing the example from \figref{openDb-authors} with subsequent dynamically-determined data.}
\end{figure}

Observe that this code alternates between \emph{dynamically determining} the
types of the rows and tables (by reading the two files \lstinline!authors.csv!
and \lstinline!books.csv! on lines~\ref{line-openDb-authors-full}
and~\ref{line-openDb-books}) and computing over those tables (on
lines~\ref{line-filterDb-full} and~\ref{line-joinDb}).
This example is a simple version of a pervasive pattern in dynamic languages,
where execution interleaves dynamic steps that create new data types
(the tables loaded in lines~\ref{line-openDb-authors-full}
and~\ref{line-openDb-books}) and steps that compute over
previously-defined data types (the filtering and joining steps in
lines~\ref{line-filterDb-full} and~\ref{line-joinDb}).

\begin{figure}[tb]\centering
\subfloat[Phaseless, online verification-validation.]{
\label{fig:phaseless-grid}
\includegraphics[width=0.45\linewidth]{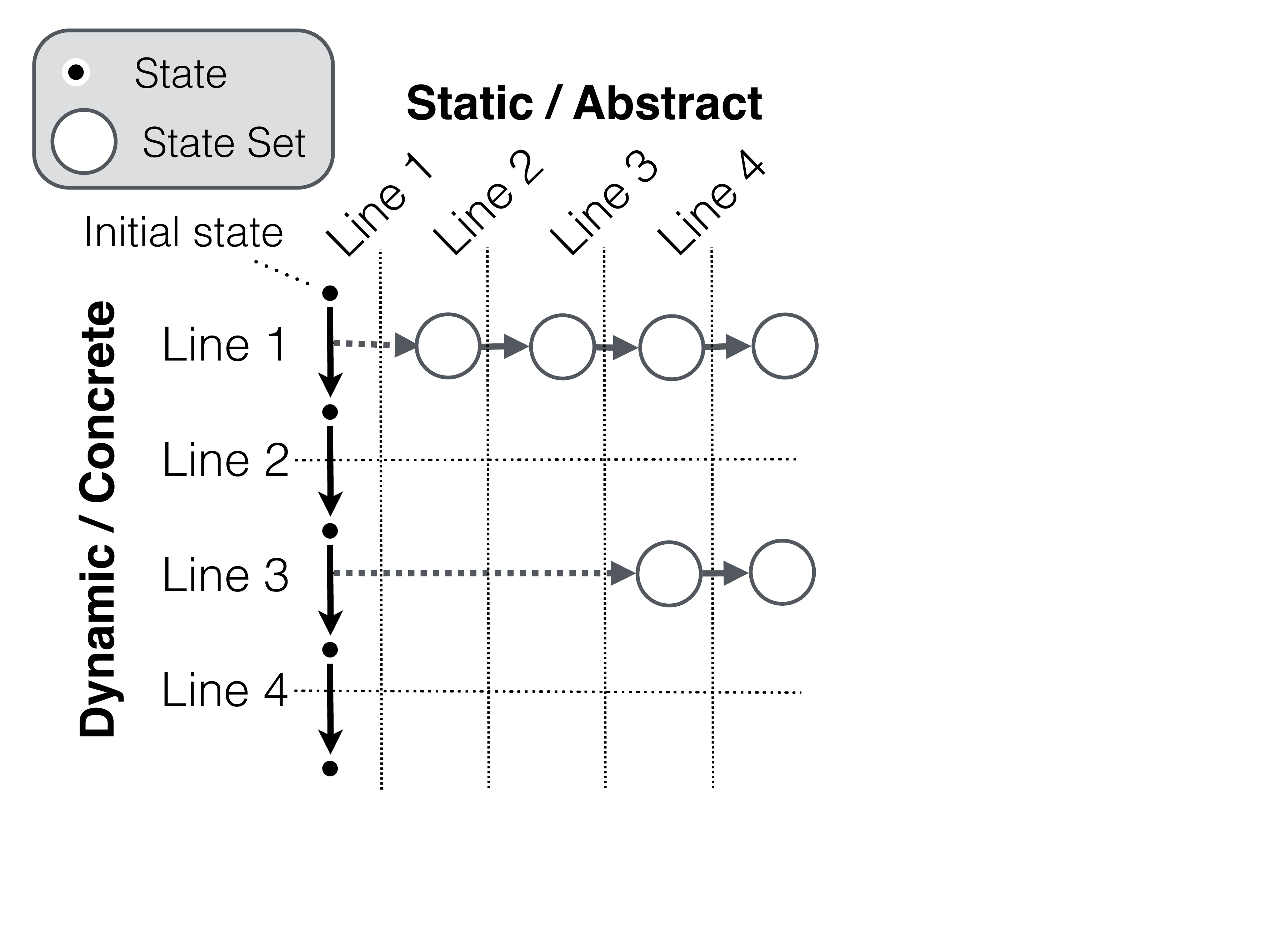}
}
\hfill\subfloat[Phasic, offline verification and online validation.]{
\label{fig:phasic-grid}
\includegraphics[width=0.45\linewidth]{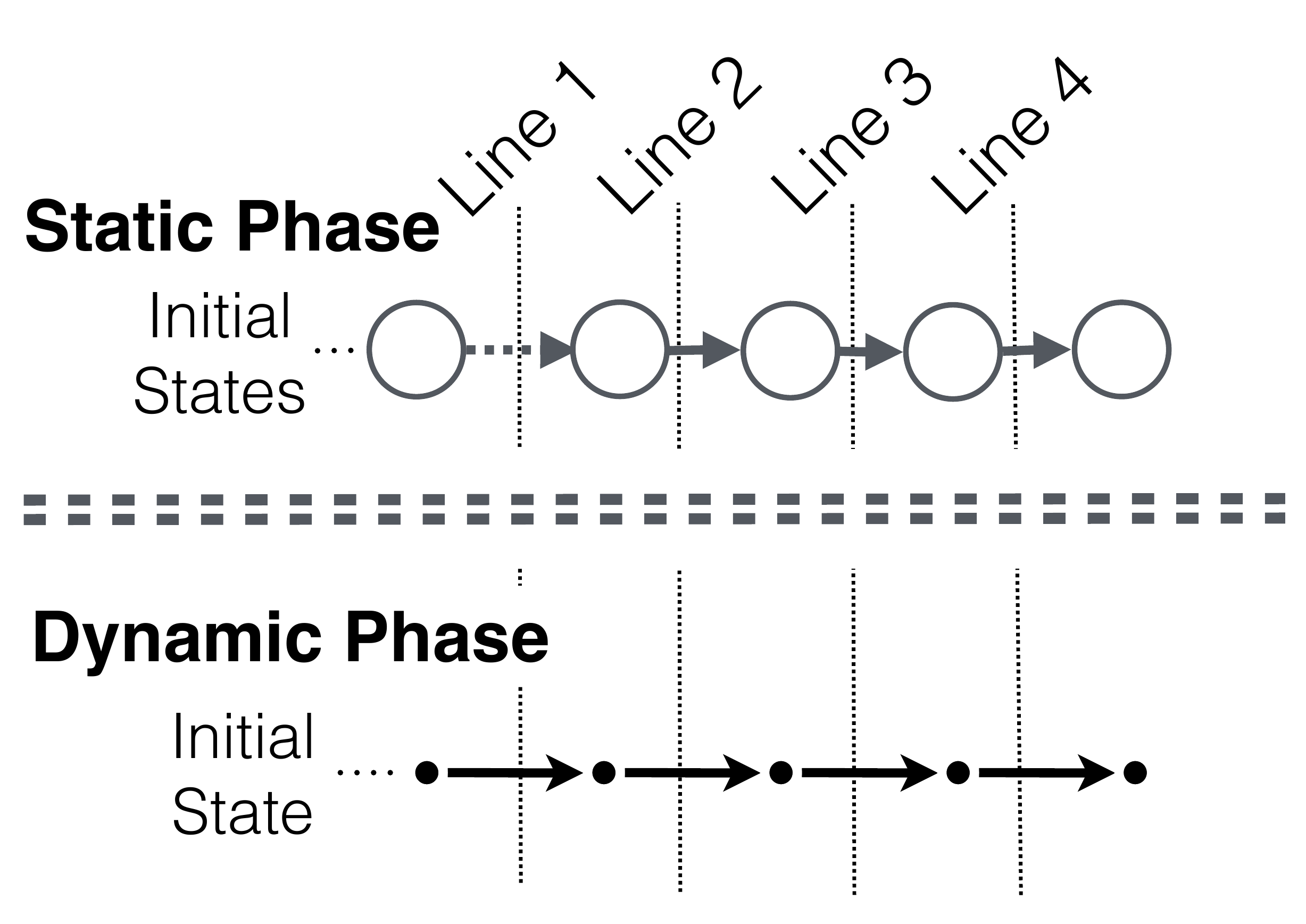}
}
\caption{Typical type checkers are phasic with
an offline static verifier and an online dynamic validator.}
\end{figure}

\paragraph{Phaseless Analysis.}
The essence of online verification-validation is pausing concrete execution to
interleave it with abstract interpretation. Pictorially, this interleaving of
concrete and abstract execution can be visualized as a two-dimensional grid, as
in \figref{phaseless-grid}.
The vertical axis represents the extent of concrete execution and dynamic,
\ExistsAnalysis (measured by program line), and for each such point, the
horizontal axis represents the extent of abstract interpretation and static,
\ForallAnalysis performed at this dynamic execution point.  As shown, abstract
interpretation explores the states after lines 2--4 after pausing at line 1, but
revisits the state after line 4 again after pausing at line 3. That is, we
imagine suspending the concrete execution of the program in
\figref{full-example-code} after the \code{openDb} call on
\reftxt{line}{line-openDb-authors-full} and then interpreting the continuation
to statically compute, under an abstraction, the set of reachable future states
from this suspended current state (the horizontal axis).

This online look-ahead would permit us to check the projections of
\code{author.citizenship} and \code{db1[i]["name"]} (on
lines~\ref{line-filterDb-full} and \ref{line-joinDb}, respectively) before the
running program executes them concretely.
However, the other projection in the \code{joinDb} call on
\reftxt{line}{line-joinDb}---the projection of the \code{author} field (i.e.,
\code{db2[j]["author"]})---cannot be proven valid in this continuation because
the \lstinline!books.csv! database has not yet been loaded. But imagine
similarly suspending the concrete execution again after
\reftxt{line}{line-openDb-books} (shown as the bottom horizontal execution in
\figref{phaseless-grid}). Now in this continuation, an abstract interpretation
can easily verify this last remaining projection on all future concrete
executions from this point.

For exposition, this example is short, and the distance between the
concrete points for static checks and the future concrete points of
potential failure (a bad projection) are tiny: they only consist of
one or two lines.
In the general case, however, the distance between these points can be
arbitrarily large.  For instance, a long-running scientific workload
may last \emph{days} or \emph{weeks}, and in these cases, it is
critical to know about possible future execution failures as soon as
possible, to minimize interruption due to programming errors.

\paragraph{Related Work: Traditional Analyses are Phasic.}
Typical, existing program analyses impose \emph{phase distinctions} between
static and dynamic steps, forcing each phase to use one approach or the other.
At one extreme, today's techniques for static verification explore all
possible execution paths, but have no knowledge of the dynamic
execution environment.
At the other extreme, dynamic validation explores one path of
execution: the path determined by concrete execution.

Type system design is, in general, a tradeoff in checking in an offline, static
phase (e.g., are function application expressions well-typed?) and checking in
the online, dynamic phase (e.g., is an array-index in bounds?). We illustrate
this phasic architecture in \figref{phasic-grid} where the static and dynamic
phases are sequenced and independent.

Some techniques attempt to ``blend'' these static and dynamic phases,
so that information gleaned from one phase feeds into the other.
For instance, gradual typing~\cite{siek+2006:gradual-typing,siek+2007:gradual-typing} enables shifting type checking between
the static and dynamic phases.
Or, the notion of using a set of dynamic runs to glean information \emph{before}
static verification appears several times in the literature for call
resolution~\cite{dufour+2007:blended-analysis}, reflection
instantiation~\cite{DBLP:conf/icse/BoddenSSOM11} and for
\lstinline{eval}~\cite{furr+2009:profile-guided-static,DBLP:conf/issta/WeiR13}.
We consider such techniques phasic if there is some sequencing (rather than
interleaving) of static, \ForallAnalysis{} and dynamic, \ExistsAnalysis{} phases.

While phasic analyses dominate the literature, there exist some
non-phasic analysis techniques, such as the ``proofs-from-tests''
approach~\cite{beckman+2010:proofs-from},
that mix \ExistsAnalysis{} information during
\ForallAnalysis{} (e.g., via directed-random automated
testing~\cite{godefroid+2005:dart:-directed}).
In these works, the goal is to perform \emph{offline} static
verification whose abstraction selection leverages \ExistsAnalysis
information from testing.
This is distinct from our vision, where concrete execution is
interposed with analyses (not vice versa).
However, this work shares our concern with incremental exploration of
a state space, and it possible that our proposed incremental substrate
would also be beneficial in the context of ``proofs-from-tests.''

\paragraph{Online Verification-Validation.}

OVV programs consist of two stratified layers that interact during execution.
First, the \emph{object layer} expresses ordinary execution. Execution can
\emph{escape} into the \emph{meta layer}, which is capable of inspecting the
run-time representation of the object layer.
Code at the meta layer expresses \ForallAnalysis{} and
\ExistsAnalysis{} over object programs.
During ordinary execution, the meta layer plays a passive role until
special primitives transfer control.  In particular, \code{reflect/cc}
transfers control, along with a reflected view of the current (object
layer) continuation.
Further, the meta layer has access to read and write hidden annotations on
object layer values (e.g., to store program facts, such as types).
Collectively, these hidden annotations can be viewed as providing a
``shadow heap'' for tracking dynamic, meta-level information, in the
service of performing \ExistsAnalyses.
For maximum extensibility, the object layer lacks a static type
system, relying on checking at the other layer. For convenience in
expressing analysis over object programs, the meta layer may employ a
language that employs a static type system (a la ML), but this is not
a requirement. For concreteness in presentation, we use JavaScript
syntax for object layer code and Rust-like syntax for meta layer code,
and we may consider language choice as an orthogonal concern.

\begin{figure}
\begin{lstlisting}[style=number]
function openDb(file) {
  var table = openDbInternal(file)
  {{ reflect/cc (chk_state, return table) }}
}
\end{lstlisting}
\caption{The \code{openDb} library function synthesizes a type for the loaded
table before returning via \code{reflect/cc}.}
\label{fig:openDb}
\end{figure}
\paragraph{\code{reflect/cc}.}
The implementor of \code{openDb} uses \code{reflect/cc} to mediate between
concrete execution in the object layer, and interposed code in the meta layer
that performs online \ForallAnalysis (shown in \figref{openDb}).
First, \code{openDb} uses \code{openDbInternal} to load the given file
and construct a table from its CSV content (line 2).
Next, line 3 uses \code{reflect/cc} to invoke the meta-level function
\code{chk_state}, which determines types for the table's content
(failing with an error if this content is malformed).
As a side effect, it writes to a meta layer field on the \code{table}
variable to record the type of the rows of the table so that it can be
consumed in a subsequent meta layer execution.
In particular, the \code{reflect/cc}~primitive pauses the execution of the
object program, giving control to a meta language function, here~\code{chk_state}.
Before transferring control, the primitive reflects the current continuation as
a \emph{first-class data structure} and uses it as an argument to the given meta
language code, here~\code{chk_state}, as we illustrate in \figref{reflectcc-grid}.

\newsavebox{\SBoxOpenDbAbbrev}
\begin{lrbox}{\SBoxOpenDbAbbrev}
\begin{minipage}[b]{0.4\linewidth}
\begin{lstlisting}[style=number,xleftmargin=0mm]
let $\ldots$ = openDb($\ldots$)
let $\ldots$ = filterDb($\ldots$)
let $\ldots$ = openDb($\ldots$)
let $\ldots$ = joinDb($\ldots$)
\end{lstlisting}
\end{minipage}
\end{lrbox}

\begin{figure}\centering
\subfloat[The transfer of control moves from the object program to
the meta program via \code{reflect/cc}.]{
\includegraphics[width=0.45\linewidth]{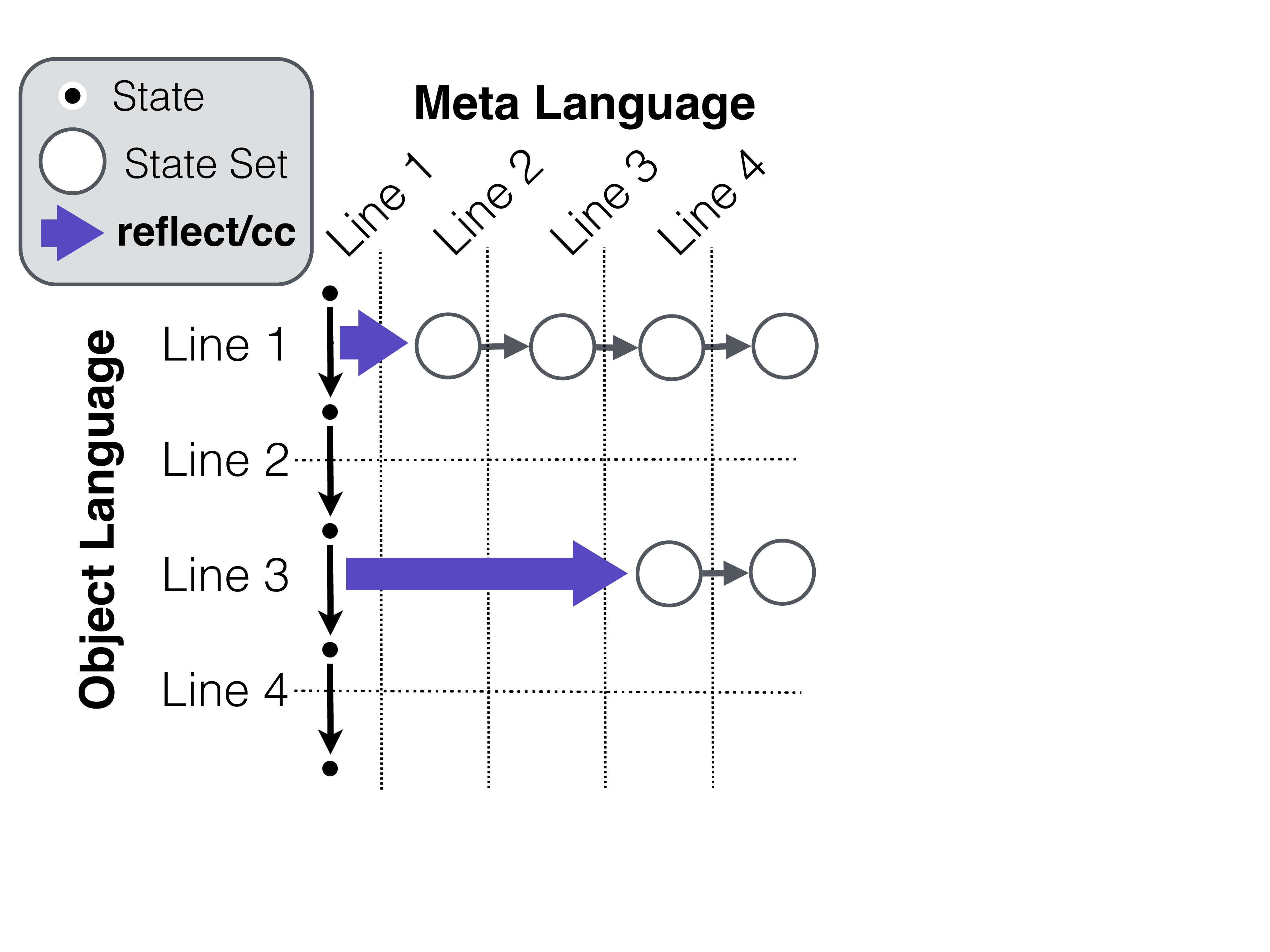}
}
\hfill\subfloat[Sketch of the example code from \figref{full-example-code}.]{
\usebox{\SBoxOpenDbAbbrev}
}
\caption{Illustrating online verification-validation.}
\label{fig:reflectcc-grid}
\end{figure}

By using this reflected structure, the meta language code can perform
arbitrarily complex \ForallAnalysis.  When it finishes, it returns
control to the object program by returning a transformed program state.
In this case, \code{openDb} uses \code{chk_stk} to type-check the
program's continuation at each call, given the type of the loaded
table.
Based on this type information, \code{openDb} either signals errors
(if there is a type error), or transforms future validation checks (if
type analysis succeeds in proving that these validation checks are
redundant).


\lstset{
  morecomment=[l]{//}
}
\begin{figure}
\begin{tabular*}{\linewidth}{@{\extracolsep{\fill}}cc@{}}
\begin{minipage}{0.45\columnwidth}
\begin{lstlisting}[language=JavaScript]
enum Stk {
  Halt,
  FrLet(Env,Var,Exp,Stk),
  FrApp(Val,Stk)
}
\end{lstlisting}
\end{minipage}
&
\begin{minipage}{0.45\columnwidth}
\begin{lstlisting}[language=JavaScript]
enum CTyp {
  Ret(VTyp),
  Arr(VTyp, CTyp)
}
\end{lstlisting}
\end{minipage}
\end{tabular*}
\begin{lstlisting}[language=JavaScript]
meta chk_stk (st:StTyp, stk:Stk, ct:CTyp) -> Option<Stk>
{ match (ct, stk) {
   _,       Halt => Some(stk),
   Ret(vt), FrLet(env, x, e, stk2) =>
   { match syn_tenv(st, env) { None => None,
     Some(tenv,env2) =>
     { let tenv = tenv_ext(tenv, x, vt) ;
       match syn_exp(st, tenv, e) { None => None,
       Some(et, e2) =>
       { match chk_stack (st, stk2, et) { None => None,
         Some(stk3) => Some(FrLet(env2, x, e2, stk3))
       }}
     }}
   }}
   Arr(vt,ct), FrApp(v,stk2) => {
     match ( check_value(st, emp, v, vt),
             check_stack(st, stk2, c) ) {
       (Some(v3), Some(stk3)) => Some(App(v3,stk3)),
       _                      => None,
   }}
}}
\end{lstlisting}
\caption{Type checking continuations.}
\label{fig:chkCont}
\end{figure}

\paragraph{The Meta Program Expresses Online Meta Theory.}
\figref{chkCont} lists the meta-layer function~\code{chk_stk}, which
checks a VM stack against a store typing and computation type.
We adopt Rust-like syntax (a recent dialect of ML).
%
In particular, this meta-level code interacts with the store typing,
stack and computation types as values, as if they are persistent
(purely functional, applicative) inductively-defined structures, a la
ordinary ML.
(The top of \figref{chkCont} defines these structures in Rust-like
syntax).
Because the meta layer has \emph{meta-level access} to the semantic
structures of the object layer, it is sufficiently powerful to compute
meta-theoretical properties (either \ForallAnalysis or
\ExistsAnalysis).
In particular, meta layer function~\code{chk_stk} performs a \ForallAnalysis
(type-inference and checking), and uses these universal program facts
to improve the efficiency of \ExistsAnalyses that will occur in the
future.

At a high level, \code{chk_stk} traverses the frames of a VM stack
(from top to bottom), and checks whether the form of the frame is
consistent with the given computation type~\code{ct}.
There are three cases to consider, depending on whether the stack is
empty (\code{Halt}), or has a top-most frame for a \code{let} binding
or function application.

The computation type~\code{Ret(vt)} indicates that the local
continuation will return a value of type~\code{vt}, and that the stack
should contain a \code{let} body that is expecting to bind this value.
The frame~\code{FrLet(env,x,e)} consists of a saved VM
environment~\code{env} (mapping local variables to values), a variable
to let-bind~\code{x}, and a body~\code{e} in which the variable is
scoped.
To check this case, the code first attempts to synthesize a typing
environment from \code{env}; if this fails, the stack does not check,
and verification fails.
Otherwise, the meta-level function~\code{syn_tenv} synthesizes a
typing environment and returns an annotated version of the given
environment,~\code{env2}.
Next, the case attempts to synthesize a type~\code{et} for the
\code{let}~body,~\code{e}.
When successful, synthesis produces a type~\code{et} and an annotated
term~\code{e}.
Finally, the case checks the recursive structure of the stack; when
successful, it returns a transformed stack, whose environments, terms
and values are annotated.

The computation type~\code{Arr(vt,ct2)} indicates that the local
continuation is a function abstraction that will consume a value of
type~\code{vt}, and that the stack should contain an argument value
with this type.
Similar to above, the case checks the value has the correct
type~\code{vt}, and checks the stack recursively; when successful, it
returns a transformed (annotated) stack.

\paragraph{Defining and using OVV.}
Below, \secref{vmf} defines the syntax and dynamic semantics of
\lambdaVMF in detail; building on these definitions, \secref{typing}
revisits the algorithm shown in \figref{chkCont}, showing it in the
context of a larger system for gradually-typing programs that compute
with databases, like the one shown in \figref{reflectcc-grid}.

\section{OVV Machine Semantics}
\label{sec:vm-semantics}
\label{sec:vmf}

We present \lambdaVMF, an abstract machine semantics for libraries
and programs that employ OVV.  In \secref{typing}, we instantiate
this OVV framework with a gradual type system for checking programs
that compute with simple databases.

\paragraph{Program Syntax.}
\figref{syntax} gives the syntax for \lambdaVMF programs.
To streamline the definition of analyses and dynamic interpretation,
\lambdaVMF syntactically separates program structure into expressions~$\ottnt{e}$
and values~$\ottnt{v}$.
Further, to permit a meta-layer, extension-defined analysis to
annotate \lambdaVMF programs (as illustrated in \secref{typing}), the
recursive syntax of expressions and values consists of annotated
pre-expressions~$\dot{e}$ and annotated pre-values~$\dot{v}$,
respectively.
The operational semantics of \lambdaVMF programs, defined below, does
not directly depend on these annotations.  However, an extension
may use its annotations to prove properties that verify sound online
program transformations.  In these cases, the annotations can
indirectly impact program behavior, i.e., by aiding a meta-level
program transformation.

Pre-values consist of open and closed thunks, which represent suspended
expressions, including all higher-order data~($\ottkw{othunk} \, \ottnt{e}$ and
$\ottkw{thunk} \, \rho \, \ottnt{e}$); closed thunks employ a closing
environment~$\rho$ that maps the free variables of~$\ottnt{e}$ to
(closed) values.
Base types consist of numbers~($\ottkw{num} \, \ottmv{n}$), strings~($\ottkw{str} \, \ottmv{s}$),
boolean bits~($\ottkw{bool} \, \ottnt{b}$) and reference cells~($\ottkw{loc} \, \ell$).
Dictionaries map values to values~($\ottkw{dict} \, \delta$), modeling a row
of a database, or a record, where the typical notion of a field name
is generalized to any value in \lambdaVMF.

Pre-expressions consist of forcing a suspended expression~($\ottkw{force} \, \ottnt{v}$),
function abstraction~($ \lambda  x .  \ottnt{e} $), function application~($\ottnt{e} \, \ottnt{v}$), let-binding a returned value ($ \ottkw{let} \; x \,{\texttt{=} }\, \ottnt{e_{{\mathrm{1}}}} \, \ottkw{in} \, \ottnt{e_{{\mathrm{2}}}} $),
returning a value~($\ottkw{ret} \, \ottnt{v}$), allocating, mutating and accessing
mutable storage ($\ottkw{ref} \, \ottnt{v}$, $\ottkw{set} \, \ottnt{v_{{\mathrm{1}}}} \, \ottnt{v_{{\mathrm{2}}}}$, $\ottkw{get} \, \ottnt{v}$,
respectively), updating the field of a record ($\ottkw{ext} \, \ottnt{v_{{\mathrm{1}}}} \, \ottnt{v_{{\mathrm{2}}}} \, \ottnt{v_{{\mathrm{3}}}}$),
projecting the field of a record ($ \ottnt{v_{{\mathrm{1}}}} \left[  \ottnt{v_{{\mathrm{2}}}}  \right]_{ a^{\textsf{op} } } $).

Finally, \lambdaVMF includes special forms for ascribing a
sub-expression with a manual annotation ($\ottnt{e}  \mathrel{ \texttt{?:} }  a^\textsf{e}$),
and reflectively inspecting (and transforming) the current continuation via
\lstinline{reflect/cc} as core primitive, $\ottkw{rcc} \, e_\textsf{m} \, \ottnt{e}$.
Notably, execution pauses \emph{before} executing the local
continuation~$\ottnt{e}$, and a common idiom consists of using a manual
ascription there, to be discharged via the use of $\ottkw{rcc}$.

As explained below, using $\ottkw{rcc}$ to prove and discharge
ascriptions is actually necessary in \lambdaVMF, since they have no
other form of dynamic semantics. \TODO{I think we should expand this discussion. This is the interesting part. -Evan}
For this purpose, the meta-level program $e_\textsf{m}$ transforms the
program state before its continuation resumes.
We do not model the model-level programming language here; our current
implementation uses Rust.

\begin{figure}
\small
\begin{jgrammar}
  Value    & $\ottnt{v}$& $\bnfas$& $ \dot{v} ~\texttt{@:}~ a^\textsf{v} $ & Annotated pre-value
  \\
  Pre-Val. & $\dot{v}$
  &$\bnfas$&        $\ottkw{othunk} \, \ottnt{e}$    & Open thunk
  \\ &&& $\bnfaltbrk \ottkw{thunk} \, \rho \, \ottnt{e}$ & Closed thunk
  \\ &&& $\bnfaltbrk \ottkw{dict} \, \delta$  & Dictionary
  \\ &&& $\bnfaltbrk \ottkw{num} \, \ottmv{n}$ & Number
  \\ &&& $\bnfaltbrk \ottkw{str} \, \ottmv{s}$ & String
  \\ &&& $\bnfaltbrk \ottkw{bool} \, \ottnt{b}$ & Boolean
  \\ &&& $\bnfaltbrk \ottkw{loc} \, \ell$ & Store location
  \\ &&& $\bnfaltbrk \ottsym{()}$    & Unit value
  \\ &&& $\bnfaltbrk x$     & Value variable
  \\
  Dict. & $\delta$
  &$\bnfas$& $\varepsilon~|~ \delta  ,  \ottnt{v_{{\mathrm{1}}}}  \mapsto  \ottnt{v_{{\mathrm{2}}}} $ & Dictionaries of values
  \\[2mm]
  Expr.     & $\ottnt{e}$& $\bnfas$& $ \dot{e} ~\texttt{@:}~ a^\textsf{e} $ & Annotated pre-expression
  \\
  Pre-Expr. & $\dot{e}$ & $\bnfas$ & $\ottkw{force} \, \ottnt{v}$ & Unsuspend (force) thunk
  \\ &&& $\bnfaltbrk  \lambda  x .  \ottnt{e} $ & Function abstraction
  \\ &&& $\bnfaltbrk \ottnt{e} \, \ottnt{v}$ & Function application
  \\ &&& $\bnfaltbrk  \ottkw{let} \; x \,{\texttt{=} }\, \ottnt{e_{{\mathrm{1}}}} \, \ottkw{in} \, \ottnt{e_{{\mathrm{2}}}} $ & Bind computed value
  \\ &&& $\bnfaltbrk \ottkw{ret} \, \ottnt{v}$ & Produce a value
  \\ &&& $\bnfaltbrk \ottkw{ref} \, \ottnt{v}$ & Allocate store reference
  \\ &&& $\bnfaltbrk \ottkw{set} \, \ottnt{v_{{\mathrm{1}}}} \, \ottnt{v_{{\mathrm{2}}}}$ & Store mutation
  \\ &&& $\bnfaltbrk \ottkw{get} \, \ottnt{v}$ & Store projection
  \\ &&& $\bnfaltbrk \ottkw{ext} \, \ottnt{v_{{\mathrm{1}}}} \, \ottnt{v_{{\mathrm{2}}}} \, \ottnt{v_{{\mathrm{3}}}}$ & Dictionary extension
  \\ &&& $\bnfaltbrk  \ottnt{v_{{\mathrm{1}}}} \left[  \ottnt{v_{{\mathrm{2}}}}  \right]_{ a^{\textsf{op} } } $ & Dictionary projection
  \\ &&& $\bnfaltbrk \ottnt{e}  \mathrel{ \texttt{?:} }  a^\textsf{e}$ & Annotation ascription
  \\ &&& $\bnfaltbrk \ottkw{rcc} \, e_\textsf{m} \, \ottnt{e}$ & Reflect current continuation
  \\[2px]
  Annot.
    & $a^{\textsf{op} }$ & $\bnfas$ & $ \textsf{?} ~|~ \texttt{!} $ & Uncertain vs. certain
  \\
    & $a^\textsf{v}$ & $\bnfas$ & $\cdots$ & Value annotation
  \\
    & $a^\textsf{e}$ & $\bnfas$ & $\cdots$ & Expression annotation
 \\[2px]
 Meta Expr.
   & $e_\textsf{m}$ & $\bnfas$ & $\cdots$ & Meta-level programs
\end{jgrammar}
\caption{Syntax of \lambdaVMF Programs}
\label{fig:syntax}
\end{figure}

\begin{figure}
  \small
  \begin{jgrammar}
    State
    & $\sigma$ & $\bnfas$ & $ \left<  \mu ;  \kappa ;  \rho ;  \dot{e}  \right> $ &
    \\
    Store
    & $\mu$ & $\bnfas$ & $\varepsilon~|~ \mu  ,  \ell \mapsto \ottnt{v} $ & Maps locations to values
    \\
    Stack
    & $\kappa$ & $\bnfas$ & $ \textsf{halt} $ & Empty stack
    \\
    &&& $\bnfaltbrk \kappa  ::  \ottsym{(}  \rho  \ottsym{,}  x  \ottsym{.}  \ottnt{e}  \ottsym{)}$ & Waiting for return
    \\
    &&& $\bnfaltbrk \kappa  ::  \ottnt{v}$ & Fun. application argument
    \\
    Environment
    & $\rho$   & $\bnfas$ & $\varepsilon~|~ \rho  ,  x \mapsto \ottnt{v} $ & Maps variables to values
  \end{jgrammar}
  \caption{VM State: The store, stack and environment.}
  \label{fig:state}
\end{figure}

\paragraph{VM State syntax.}
\figref{state} defines the global state of the \lambdaVMF program: It
consists of a store, mapping locations to mutable
values~($\mu$); a stack of evaluation context frames
($\kappa$), an environment mapping variables to values~($\rho$),
and a pre-expression~$\dot{e}$ that gives the current local
continuation.
Non-empty stacks give evaluation contexts for $\ottkw{let}$
bodies~($\kappa  ::  \ottsym{(}  \rho  \ottsym{,}  x  \ottsym{.}  \ottnt{e}  \ottsym{)}$) and function
application~($\kappa  ::  \ottnt{v}$).

\paragraph{Dynamics of \lambdaVMF Programs.}

\begin{figure*}
\small
\judgbox{ \sigma  \longrightarrow  \sigma' }{
  \lambdaVMF state~$\sigma$ steps to state~$\sigma'$.
}
\[
\begin{array}{@{\hspace{-10mm}}rlclcl}
  \left<  \mu ;  \kappa ;  \rho  \right. & \left.   \ottkw{let} \; x \,{\texttt{=} }\, \ottsym{(}   \dot{e}_{{\mathrm{1}}} ~\texttt{@:}~  \_    \ottsym{)} \, \ottkw{in} \, \ottnt{e_{{\mathrm{2}}}}   \right>   & \longrightarrow &   \left<  \mu ;  \kappa  ::  \ottsym{(}  \rho  \ottsym{,}  x  \ottsym{.}  \ottnt{e_{{\mathrm{2}}}}  \ottsym{)} ;  \rho ;  \dot{e}_{{\mathrm{1}}}  \right>  
\\
  \left<  \mu ;  \kappa ;  \rho  \right. & \left.  \ottsym{(}   \dot{e} ~\texttt{@:}~  \_    \ottsym{)} \, \ottnt{v}  \right>   & \longrightarrow &   \left<  \mu ;  \kappa  ::   \rho ( \ottnt{v} )  ;  \rho ;  \dot{e}  \right>  
\\
  \left<  \mu ;  \kappa  ::  \ottsym{(}  \rho  \ottsym{,}  x  \ottsym{.}  \ottsym{(}   \dot{e} ~\texttt{@:}~  \_    \ottsym{)}  \ottsym{)} ;  \rho'  \right. & \left.  \ottkw{ret} \, \ottnt{v}  \right>   & \longrightarrow &   \left<  \mu ;  \kappa ;   \rho  ,  x \mapsto  \rho ( \ottnt{v} )   ;  \dot{e}  \right>  
\\
  \left<  \mu ;  \kappa  ::  \ottnt{v} ;  \rho  \right. & \left.   \lambda  x .  \ottsym{(}   \dot{e} ~\texttt{@:}~  \_    \ottsym{)}   \right>   & \longrightarrow &   \left<  \mu ;  \kappa ;   \rho  ,  x \mapsto \ottnt{v}  ;  \dot{e}  \right>  
\\
  \left<  \mu ;  \kappa ;  \rho  \right. & \left.  \ottkw{force} \, \ottnt{v}  \right>   & \longrightarrow &   \left<  \mu ;  \kappa ;  \rho' ;  \dot{e}  \right>  
& \textrm{when} &  \rho ( \ottnt{v} )\,{=}\, \ottsym{(}   \ottkw{thunk} \, \rho' \, \ottsym{(}   \dot{e} ~\texttt{@:}~  \_    \ottsym{)} ~\texttt{@:}~  \_    \ottsym{)} 
\\
  \left<  \mu ;  \kappa ;  \rho  \right. & \left.  \ottkw{ref} \, \ottnt{v}  \right>   & \longrightarrow &   \left<   \mu  ,  \ell \mapsto \ottnt{v}  ;  \kappa ;  \rho ;  \ottkw{ret} \, \ottsym{(}   \ottkw{loc} \, \ell ~\texttt{@:}~ \textsf{?}   \ottsym{)}  \right>  
& \textrm{when} &  \ell  \not\in  \mu 
\\
  \left<  \mu ;  \kappa ;  \rho  \right. & \left.  \ottkw{set} \, \ottnt{v_{{\mathrm{1}}}} \, \ottnt{v_{{\mathrm{2}}}}  \right>   & \longrightarrow &   \left<   \mu  ,  \ell \mapsto  \rho ( \ottnt{v_{{\mathrm{2}}}} )   ;  \kappa ;  \rho ;  \ottkw{ret} \, \ottsym{(}   \ottsym{()} ~\texttt{@:}~ \textsf{?}   \ottsym{)}  \right>  
& \textrm{when} &  \rho ( \ottnt{v_{{\mathrm{1}}}} )\,{=}\, \ottsym{(}   \ottkw{loc} \, \ell ~\texttt{@:}~  \_    \ottsym{)} 
\\
  \left<   \mu  ,  \ell \mapsto \ottnt{v_{{\mathrm{2}}}}  ;  \kappa ;  \rho  \right. & \left.  \ottkw{get} \, \ottnt{v_{{\mathrm{1}}}}  \right>   & \longrightarrow &   \left<   \mu  ,  \ell \mapsto \ottnt{v_{{\mathrm{2}}}}  ;  \kappa ;  \rho ;  \ottkw{ret} \, \ottnt{v_{{\mathrm{2}}}}  \right>  
& \textrm{when} &  \rho ( \ottnt{v_{{\mathrm{1}}}} )\,{=}\, \ottsym{(}   \ottkw{loc} \, \ell ~\texttt{@:}~  \_    \ottsym{)} 
\\
  \left<  \mu ;  \kappa ;  \rho  \right. & \left.  \ottkw{ext} \, \ottnt{v_{{\mathrm{1}}}} \, \ottnt{v_{{\mathrm{2}}}} \, \ottnt{v_{{\mathrm{3}}}}  \right>   & \longrightarrow &   \left<  \mu ;  \kappa ;  \rho ;  \ottkw{ret} \, \ottsym{(}   \ottkw{dict} \, \ottsym{(}   \delta  ,   \rho ( \ottnt{v_{{\mathrm{2}}}} )   \mapsto   \rho ( \ottnt{v_{{\mathrm{3}}}} )    \ottsym{)} ~\texttt{@:}~ \textsf{?}   \ottsym{)}  \right>  
& \textrm{when} &  \rho ( \ottnt{v_{{\mathrm{1}}}} )\,{=}\, \ottsym{(}   \ottkw{dict} \, \delta ~\texttt{@:}~  \_    \ottsym{)} 
\\
  \left<  \mu ;  \kappa ;  \rho  \right. & \left.   \ottnt{v_{{\mathrm{1}}}} \left[  \ottnt{v_{{\mathrm{2}}}}  \right]_{  \texttt{!}  }   \right>   & \longrightarrow &   \left<  \mu ;  \kappa ;  \rho ;  \ottkw{ret} \, \ottnt{v_{{\mathrm{3}}}}  \right>  
& \textrm{when} &  \rho ( \ottnt{v_{{\mathrm{1}}}} )\,{=}\, \ottsym{(}   \ottkw{dict} \, \ottsym{(}   \delta  ,   \rho ( \ottnt{v_{{\mathrm{2}}}} )   \mapsto  \ottnt{v_{{\mathrm{3}}}}   \ottsym{)} ~\texttt{@:}~  \_    \ottsym{)} 
\\[2mm]
 \left<  \mu ;  \kappa ;  \rho  \right. & \left.   \ottnt{v_{{\mathrm{1}}}} \left[  \ottnt{v_{{\mathrm{2}}}}  \right]_{  \texttt{?}  }   \right>  & \multicolumn{2}{l}{\textit{no stepping rule}}
\\
 \left<  \mu ;  \kappa ;  \rho  \right. & \left.  \ottnt{e}  \mathrel{ \texttt{?:} }  a^\textsf{e}  \right>  & \multicolumn{2}{l}{\textit{no stepping rule}}
\\[2mm]
  \left<  \mu ;  \kappa ;  \rho  \right. & \left.  \ottkw{rcc} \, e_\textsf{m} \, \ottsym{(}   \dot{e} ~\texttt{@:}~  \_    \ottsym{)}  \right>   & \longrightarrow &  \sigma' 
& \textrm{when} &  e_\textsf{m}  ([\![   \left<  \mu ;  \kappa ;  \rho ;  \dot{e}  \right>   ]\!]) \Downarrow_{\textsf{meta} } [\![  \sigma'  ]\!] 
\end{array}
\]
\caption{Small-step, abstract machine semantics of \lambdaVMF}
\label{fig:step}
\end{figure*}

\figref{step} defines a small-step operational semantics over \lambdaVMF states.
The rules for $\ottkw{let}$ and function application each push the
stack with a frame that is eliminated by the rules for $\ottkw{ret}$
and function abstraction, respectively.
In both cases, eliminating the frame consists of binding a value to a
variable, and continuing the program.
Forcing a thunk consists of unpacking its environment and expression,
and continuing execution with them.
Rules for allocating, mutating and accessing a reference cell in the
store are each standard.
Dictionary extension adds a field to a (possibly empty) dictionary;
and dictionary projection selects a given field's associated value,
returning it.

What makes \lambdaVMF particularly interesting is that
there are no stepping rules for $ \textsf{?} $-mode field projection
or for ascription.
To avoid getting stuck at these operations, these operations should be
either verified progressively or validated, perhaps immediately before
executing, when all relevant information is available.
To do so, the program uses $\ottkw{rcc} \, e_\textsf{m} \, \ottnt{e}$, which runs the meta-level
program $e_\textsf{m}$ on a reflected version of the current VM state: the
rule constructs the current continuation, reflects this program state
into a data structure, runs the meta-level term~$e_\textsf{m}$, and then
injects the resulting program state into a transformed
continuation~$\sigma'$.
In OVV, this step sometimes verifies and validates operations before
the program attempts to execute them, transforming the program to
remove or modify them.
In contrast to traditional phasic static verification or dynamic validation (cf.
\figref{phasic-grid}), this step does not need to be either eagerly before the
entire execution or lazily just before the potentially faulting operation. By
choosing the placement of $\ottkw{rcc}$, the program can choose how eager or
lazy the checking should be anywhere between these two extremes.
We give a detailed example below.

\section{Gradual Typing for Simple Databases}
\label{sec:typing}

In this section we present a gradual type system for \lambdaVMF and
\textsf{libDb}, an extension that permits us to express the motivating
example from \figref{full-example-code}.
We tour the type system and illustrate how OVV progressively types and
validates the operations in this example.

\begin{figure}
  \small
  \begin{jgrammar}
  & $e$
    &$\bnfas$& $\cdots$ & Existing forms (\Figref{syntax})
    \\ &&& $\bnfaltbrk  \texttt{openDb}_{ a^{\textsf{op} } }~ \ottnt{v} $ & Open database by file path
    \\ &&& $\bnfaltbrk  \texttt{filterDb}_{ a^{\textsf{op} } }~ \ottnt{v_{{\mathrm{1}}}} ~ \ottnt{v_{{\mathrm{2}}}} $ & Filter DB by predicate
    \\ &&& $\bnfaltbrk  \texttt{joinDb}_{ a^{\textsf{op} } }~ \ottnt{v_{{\mathrm{1}}}} ~ \ottnt{v_{{\mathrm{2}}}} ~ \ottnt{v_{{\mathrm{3}}}} ~ \ottnt{v_{{\mathrm{4}}}} $ & Join DBs using keys' value
\end{jgrammar}
\caption{Database Library Forms}
\label{fig:db-syntax}
\end{figure}

\paragraph{Syntax for \textsf{libDb}.}
\figref{db-syntax} extends the syntax from \figref{syntax} with the three
operations implemented by \libDb, for opening databases, filtering
them with a predicate, and joining them using named fields.
Each operation is parameterized by one or more argument values, and an
operation annotation $a^{\textsf{op} }$ that determines how to type operation.
As with record field projection, this annotation determines whether
the operation's pre-conditions for success have been fully verified
via OVV.
The gradual type system for \libDb presented here uses different rules
for certain ($ \texttt{!} $) versus uncertain ($ \textsf{?} $) reasoning modes.

\paragraph{Types for \lambdaVMF and \textsf{libDb}.}
\figref{types} instantiates the \lambdaVMF framework for a gradual
type system.  This system has (bidirectional, algorithmic) rules to
reason about \lambdaVMF code as well as the \libDb extension.
Value types consist of types for thunked computations~($\ottkw{U} \, \ottnt{C}$)
\footnote{ We follow conventions from the literature on
  call-by-push-value (CBPV) in our type syntax for thunked and
  value-returning computations, which uses special letters \textbf{U}
  and \textbf{F}, respectively~\cite{Levy99subsuming,levy2003call}.
},
dictionaries ($\ottkw{Dict} \, \Delta$, where $\Delta$ maps field values to
field types), numbers~($\ottkw{Num}$), strings~($\ottkw{Str}$), booleans~($\ottkw{Bool}$), reference
cells~($\ottkw{Ref} \, \ottnt{A}$), unit~($\ottsym{1}$), unknown~($ \textsf{?} $) and
databases ($\ottkw{Db} \, \ottnt{A}$).
Computation types consist of the arrow type for functions~($\ottnt{A}  \rightarrow  \ottnt{C}$) and value types for value-producing computations~($\ottkw{F} \, \ottnt{A}$).

\begin{figure}
\small
\begin{jgrammar}
  Annotations
    & $a^\textsf{v}$ & $\bnfas$ & $\ottnt{A}$ & Value annotation
  \\
    & $a^\textsf{e}$ & $\bnfas$ & $\ottnt{C}$ & Expression annotation
  \\
  Value Types  & $\ottnt{A}, \ottnt{B}$& $\bnfas$&
  $ \ottkw{U} \, \ottnt{C}$& Thunked computation
  \\ &&& $\bnfaltbrk \ottkw{Dict} \, \Delta$& Dictionary
  \\ &&& $\bnfaltbrk \ottkw{Num}$& Number
  \\ &&& $\bnfaltbrk \ottkw{Str}$& String
  \\ &&& $\bnfaltbrk \ottkw{Bool}$& Boolean
  \\ &&& $\bnfaltbrk \ottkw{Ref} \, \ottnt{A}$& Reference cell
  \\ &&& $\bnfaltbrk \ottsym{1}$& Unit
  \\ &&& $\bnfaltbrk  \textsf{?} $& Unknown value type
  \\ &&& $\bnfaltbrk \ottkw{Db} \, \ottnt{A}$& Database; multiset of $\ottnt{A}$s
  \\[2mm]
  Dictionary & $\Delta$ & $\bnfas$ & \multicolumn{2}{l}{$\varepsilon~|~ \Delta ,  \ottnt{v} \mapsto \ottnt{A} $~~~\textrm{Maps values to types}}
  \\[2mm]
  Computation & $\ottnt{C}, \ottnt{D}$ & $\bnfas$&
  $\ottnt{A}  \rightarrow  \ottnt{C}$ & Function abstraction
  \\
  Types
  &&& $\bnfaltbrk \ottkw{F} \, \ottnt{A}$ & Value production
\end{jgrammar}
\caption{Type Syntax: Annotations for Values and Expressions}
\label{fig:types}
\end{figure}

\begin{figure}
\small
\judgbox{\sigma \, {\textsf{ok} }}{
  State~$\sigma$ is well-typed.
}
\begin{mathpar}
\Infer{state}
{
 \left| \mu \right|   \vdash  \rho  \Rightarrow  \Gamma
\\
 \left| \mu \right|   \ottsym{,}  \Gamma  \vdash  \dot{e}  \Rightarrow  \ottnt{C}
\\
 \left| \mu \right|   \vdash  \kappa  \Leftarrow  \ottnt{C}
}
{
 \left<  \mu ;  \kappa ;  \rho ;  \dot{e}  \right>  \, {\textsf{ok} }
}
\end{mathpar}
\judgbox{\Gamma  \vdash  \kappa  \Leftarrow  \ottnt{C}}{
  Under~$\Gamma$,
  stack $\kappa$
  eliminates a computation of type~$\ottnt{C}$.
}
\begin{mathpar}
\Infer{k-emp}
{ }
{\Gamma  \vdash  \textsf{halt}  \Leftarrow  \ottnt{C}}
\and
\Infer{k-let}
{
\Gamma_{{\mathrm{1}}}  \vdash  \rho  \Rightarrow  \Gamma_{{\mathrm{2}}}
\\\\
\Gamma_{{\mathrm{2}}}  \ottsym{,}  x  \ottsym{:}  \ottnt{A}  \vdash  \ottnt{e}  \Rightarrow  \ottnt{C}
\\\\
\Gamma_{{\mathrm{1}}}  \vdash  \kappa  \Leftarrow  \ottnt{C}
}
{\Gamma_{{\mathrm{1}}}  \vdash  \kappa  ::  \ottsym{(}  \rho  \ottsym{,}  x  \ottsym{.}  \ottnt{e}  \ottsym{)}  \Leftarrow  \ottkw{F} \, \ottnt{A}}
\and
\Infer{k-app}
{
\Gamma  \vdash  \ottnt{v}  \Leftarrow  \ottnt{A}
\\\\
\Gamma  \vdash  \kappa  \Leftarrow  \ottnt{C}
}
{\Gamma  \vdash  \kappa  ::  \ottnt{v}  \Leftarrow  \ottnt{A}  \rightarrow  \ottnt{C}}
\end{mathpar}
\caption{Stack typing and State typing}
\label{fig:stack-state-typing}
\end{figure}

\paragraph{Typing \lambdaVMF program states.}

\figref{stack-state-typing} lists typing judgement forms for
\lambdaVMF program states and for stacks.
To type a program state, we assume that the stored values are
annotated, and we use these annotations as a store typing, written
$ \left| \mu \right| $, which maps reference locations to values types.
To type a program state, we assume this store typing~$ \left| \mu \right| $ and attempt to verify the the other VM machinery,
consisting of the current environment~$\rho$, program
term~$\dot{e}$, and stack~$\kappa$.
Three judgements compute type properties for these components:
Assuming a store typing $\Gamma$, the judgement $\Gamma  \vdash  \rho  \Rightarrow  \Gamma'$ computes from an environment (mapping variables to
values), a typing context~$\Gamma'$ (mapping variables to types).
Assuming a typing context $\Gamma$, the judgement $\Gamma  \vdash  \ottnt{e}  \Rightarrow  \ottnt{C}$
computes a type from a term $e$.
Assuming a typing context $\Gamma$ and computation type~$C$ for a
terminal computation, the judgement $\Gamma  \vdash  \kappa  \Leftarrow  \ottnt{C}$ checks that
the stack~$\kappa$ either correctly continues execution or halts.

The remainder of the figure gives three rules for type-checking the
stack.
First, \textrm{k-emp} says that halting stacks are always permitted.
Next, \textrm{k-let} and \textrm{k-app} handle the recursive cases of
the stack, where the topmost frame can be viewed as eliminating the
terminal computation type, call it~$\ottnt{D}$.
In the case of \textrm{k-let}, we have that~$\ottnt{D}$ is~$\ottkw{F} \, \ottnt{A}$,
which types the terminal computation that returns a value of
type~$\ottnt{A}$; we check that the top of the stack holds the body of the
let, which can use the let-bound variable (of type~$\ottnt{A}$), for which
we can synthesize another computation type~$\ottnt{C}$ that checks against
the rest of the stack.
In the case of \textrm{k-app}, we have that $\ottnt{D}$ is~$\ottnt{A}  \rightarrow  \ottnt{C}$,
the type of a function abstraction; we check that the top of the stack
is an argument value of type~$\ottnt{A}$, and the rest of the stack checks
against the type of the abstraction's body, $\ottnt{C}$.

\paragraph{Transforming \lambdaVMF program states.}
\JEDI{Instead of merely checking them, we transform program states}
Though the gradual type system defined here is stated
propositionally, it constitutes an algorithm, and we demonstrate this
fact by implementing these relational definitions as a (mutually)
recursive total functions.
However, instead of merely returning \texttt{true} or \texttt{false}
to indicate the success or failure of the relation to hold, in the
case of \texttt{true}, we also construct an annotated term, possibly
with transformations (e.g., changing operation annotations from
uncertain~$ \textsf{?} $ to certain~$ \texttt{!} $).

For instance, we implement the type relation for program states as a
total function from program states to (optional) program states with
annotations; and when the algorithm fails, it returns \texttt{None}:
Furthermore, this algorithm plays the role of~$e_\textsf{m}$ in
\textsf{libDb}'s use of~$\ottkw{rcc} \, e_\textsf{m} \, \ottnt{e}$.

The ability to phrase the typing relations as functional algorithms
stems the fact that the rules treat certain positions of their
(bidirectional) relations consistently as inputs and outputs, and that
outputs are determined functionally from inputs.
As an example, \figref{chkCont} gives the algorithmic version of the
stack-checking relation~$\Gamma  \vdash  \kappa  \Leftarrow  \ottnt{C}$, which resembles an
ordinary function in ML.
The remainder of the rules transform in a similar manner, so that
checking relations produce an optional, annotated term structure,
while synthesizing relations produce an optional pair of annotated
term structure and synthesized type.
When these functions produce \code{None}, the corresponding typing
relation is not derivable.
The dynamic semantics of \lambdaVMF do not permit execution to
continue when this occurs.


\begin{figure}
\small


\judgbox{\Gamma  \vdash  \ottnt{e}  \Leftarrow  \ottnt{C}}{ Under $\Gamma$, expression $\ottnt{e}$ checks against type~$\ottnt{C}$. }
\begin{mathpar}
\Infer{sub}
{\Gamma  \vdash  \ottnt{e}  \Rightarrow  \ottnt{C}
  \\\\
   \ottnt{C}  \approx_{\textsf{?} }  \ottnt{D} }
{\Gamma  \vdash  \ottnt{e}  \Leftarrow  \ottnt{D}}
\and
\Infer{lam}
{\Gamma  \ottsym{,}  x  \ottsym{:}  \ottnt{A}  \vdash  \ottnt{e}  \Leftarrow  \ottnt{C}}
{\Gamma  \vdash   \lambda  x .  \ottnt{e}   \Leftarrow  \ottnt{A}  \rightarrow  \ottnt{C} }
\end{mathpar}
\judgbox{\Gamma  \vdash  \ottnt{e}  \Rightarrow  \ottnt{C}}{ Under $\Gamma$, expression $\ottnt{e}$ synthesizes type~$\ottnt{C}$. }
\begin{mathpar}
\Infer{annot}
{\Gamma  \vdash  \ottnt{e}  \Leftarrow  \ottnt{C}}
{\Gamma  \vdash  \ottnt{e}  \mathrel{ \texttt{?:} }  \ottnt{C}  \Rightarrow  \ottnt{C}}
\and
\Infer{app}
{\Gamma  \vdash  \ottnt{e}  \Rightarrow  \ottnt{A}  \rightarrow  \ottnt{C}
\\\\
\Gamma  \vdash  \ottnt{v}  \Leftarrow  \ottnt{A}
}
{\Gamma  \vdash  \ottnt{e} \, \ottnt{v}  \Rightarrow  \ottnt{C}}
\and
\Infer{p?}
{\Gamma  \vdash  \ottnt{v_{{\mathrm{1}}}}  \Rightarrow  \textsf{?}
\\\\
\Gamma  \vdash  \ottnt{v_{{\mathrm{2}}}}  \Rightarrow  \ottnt{B}
}
{\Gamma  \vdash   \ottnt{v_{{\mathrm{1}}}} \left[  \ottnt{v_{{\mathrm{2}}}}  \right]_{  \texttt{?}  }   \Rightarrow  \ottkw{F} \, \textsf{?} }
\and
\Infer{p!}
{\Gamma  \vdash  \ottnt{v_{{\mathrm{1}}}}  \Rightarrow  \ottkw{Dict} \, \ottsym{(}   \Delta ,  \ottnt{v_{{\mathrm{2}}}} \mapsto \ottnt{A}   \ottsym{)}
\\\\
\Gamma  \vdash  \ottnt{v_{{\mathrm{2}}}}  \Rightarrow  \ottnt{B}
}
{\Gamma  \vdash   \ottnt{v_{{\mathrm{1}}}} \left[  \ottnt{v_{{\mathrm{2}}}}  \right]_{  \texttt{!}  }   \Rightarrow  \ottkw{F} \, \ottnt{A} }
\and
\Infer{openDb?}
{\Gamma  \vdash  \ottnt{v}  \Rightarrow  \ottkw{Str}}
{\Gamma  \vdash   \texttt{openDb}_{  \texttt{?}  }~ \ottnt{v}   \Rightarrow  \ottkw{F} \, \ottsym{(}  \ottkw{Db} \, \textsf{?}  \ottsym{)}}
\and
\Infer{filterDb?}
{\Gamma  \vdash  \ottnt{v_{{\mathrm{1}}}}  \Rightarrow  \ottkw{Db} \, \textsf{?}
\\\\
\Gamma  \vdash  \ottnt{v_{{\mathrm{2}}}}  \Leftarrow  \ottkw{U} \, \ottsym{(}  \textsf{?}  \rightarrow  \ottkw{F} \, \ottkw{Bool}  \ottsym{)}
}
{\Gamma  \vdash   \texttt{filterDb}_{  \texttt{?}  }~ \ottnt{v_{{\mathrm{1}}}} ~ \ottnt{v_{{\mathrm{2}}}}   \Rightarrow  \ottkw{F} \, \ottsym{(}  \ottkw{Db} \, \textsf{?}  \ottsym{)}}
\and
\Infer{filterDb!}
{\Gamma  \vdash  \ottnt{v_{{\mathrm{1}}}}  \Rightarrow  \ottkw{Db} \, \ottnt{A} \\  \textsf{?}  \notin  \ottnt{A} 
\\\\
\Gamma  \vdash  \ottnt{v_{{\mathrm{2}}}}  \Leftarrow  \ottkw{U} \, \ottsym{(}  \ottnt{A}  \rightarrow  \ottkw{F} \, \ottkw{Bool}  \ottsym{)}
}
{\Gamma  \vdash   \texttt{filterDb}_{  \texttt{!}  }~ \ottnt{v_{{\mathrm{1}}}} ~ \ottnt{v_{{\mathrm{2}}}}   \Rightarrow  \ottkw{F} \, \ottsym{(}  \ottkw{Db} \, \ottnt{A}  \ottsym{)}}
\and
\Infer{joinDb?}
{\Gamma  \vdash  \ottnt{v_{{\mathrm{1}}}}  \Rightarrow  \ottkw{Db} \, \textsf{?}
\\
\Gamma  \vdash  \ottnt{v_{{\mathrm{2}}}}  \Rightarrow  \ottnt{B_{{\mathrm{2}}}}
\\\\
\Gamma  \vdash  \ottnt{v_{{\mathrm{3}}}}  \Rightarrow  \ottkw{Db} \, \textsf{?}
\\
\Gamma  \vdash  \ottnt{v_{{\mathrm{4}}}}  \Rightarrow  \ottnt{B_{{\mathrm{4}}}}
}
{\Gamma  \vdash   \texttt{joinDb}_{  \texttt{?}  }~ \ottnt{v_{{\mathrm{1}}}} ~ \ottnt{v_{{\mathrm{2}}}} ~ \ottnt{v_{{\mathrm{3}}}} ~ \ottnt{v_{{\mathrm{4}}}}   \Rightarrow  \ottkw{F} \, \ottsym{(}  \ottkw{Db} \, \textsf{?}  \ottsym{)}}
\and
\Infer{joinDb!}
{
\Delta  \ottsym{=}    \Delta_{{\mathrm{1}}} ,  \ottnt{v_{{\mathrm{2}}}} \mapsto \ottnt{A}   ,   \Delta_{{\mathrm{3}}} ,  \ottnt{v_{{\mathrm{4}}}} \mapsto \ottnt{A}  
\\\\
\Gamma  \vdash  \ottnt{v_{{\mathrm{1}}}}  \Rightarrow  \ottkw{Db} \, \ottsym{(}  \ottkw{Dict} \, \ottsym{(}   \Delta_{{\mathrm{1}}} ,  \ottnt{v_{{\mathrm{2}}}} \mapsto \ottnt{A}   \ottsym{)}  \ottsym{)}
\\
\Gamma  \vdash  \ottnt{v_{{\mathrm{2}}}}  \Rightarrow  \ottnt{B_{{\mathrm{2}}}}
\\\\
\Gamma  \vdash  \ottnt{v_{{\mathrm{3}}}}  \Rightarrow  \ottkw{Db} \, \ottsym{(}  \ottkw{Dict} \, \ottsym{(}   \Delta_{{\mathrm{3}}} ,  \ottnt{v_{{\mathrm{4}}}} \mapsto \ottnt{A}   \ottsym{)}  \ottsym{)}
\\
\Gamma  \vdash  \ottnt{v_{{\mathrm{4}}}}  \Rightarrow  \ottnt{B_{{\mathrm{4}}}}
}
{\Gamma  \vdash   \texttt{joinDb}_{  \texttt{!}  }~ \ottnt{v_{{\mathrm{1}}}} ~ \ottnt{v_{{\mathrm{2}}}} ~ \ottnt{v_{{\mathrm{3}}}} ~ \ottnt{v_{{\mathrm{4}}}}   \ottsym{:}  \ottkw{F} \, \ottsym{(}  \ottkw{Db} \, \ottsym{(}  \ottkw{Dict} \, \Delta  \ottsym{)}  \ottsym{)}}
\end{mathpar}
\caption{Selected typing rules for computation typing (checking and synthesis).}
\label{fig:type-terms}

\vspace*{5mm}

\begin{center}
\begin{tabular}{|l|cccc||c|}
\multicolumn{6}{c}{\textbf{Typing} (horizontal) across \textbf{Execution} (vertical). }
\\
\hline
       & Line 1      & Line 2       & Line 3        &  Line 4  & $\cdots$
\\
       & openDb      & filterDb     & openDb        & joinDb   & ---
\\
\hline
1: openDb & $ \textsf{?} $ & $ \textsf{?} $ & $ \textsf{?} $ & $ \textsf{?} $      & $ \textsf{?} $
\\
\hdashline
2: filterDb & ---     & $ \texttt{!} $ & $ \textsf{?} $ & $ \textsf{?} $      & $ \texttt{!} $ / $ \textsf{?} $
\\
3: openDb & ---     & ---     & $ \textsf{?} $ & $ \textsf{?} $      & $ \texttt{!} $ / $ \textsf{?} $
\\
\hdashline
4: joinDb & ---     & ---     & ---     & $ \texttt{!} $      & $ \texttt{!} $
\\
\hline
\end{tabular}
\end{center}
\caption{
  Progressive typing for \figref{full-example-code}:
  As execution progresses (vertically) over Lines 1 and 3, the
  continuation's typing becomes more certain: After Line 1
  executes, Line 2 types in the certain modality ($ \texttt{!} $) instead of
  the uncertain modality ($ \textsf{?} $); similarly,
  after Line 3, Lines 4 onward type using the certain modality.
  (The dashed horizontal lines indicate these progressions).
}
\label{fig:time2d-typing}

\end{figure}

\paragraph{Typing core \lambdaVMF terms bidirectionally.}
For simplicity, we use a bidirectional type system to encode the
gradual type systems of the \lambdaVMF core calculus and its
\textsf{libDb} extension.
\figref{type-terms} defines type checking (above) and synthesis
(below) for program terms.
For space reasons, we elide some synthesis cases, as well as the
checking and synthesis judgements for value forms; the rules shown
give a representative flavor for the complete definition.

The analytical (checking) judgement form $\Gamma  \vdash  \ottnt{e}  \Leftarrow  \ottnt{C}$ can be
read as, ``Under typing context $\Gamma$, term $\ottnt{e}$ checks against
computation type $\ottnt{C}$.''  Specifically, the type~$\ottnt{C}$ is given
as an input to the checking judgement, when viewed as an algorithm.
The synthesizing judgement form $\Gamma  \vdash  \ottnt{e}  \Rightarrow  \ottnt{C}$ can be read as,
``the typing context $\Gamma$ and term $\ottnt{e}$ \emph{synthesize} the
computation type $\ottnt{C}$.'' Specifically, the algorithm
\emph{computes} the type~$\ottnt{C}$, when given $\Gamma$ and $\ottnt{e}$.
For the core forms of \lambdaVMF, the bidirectional rules for values
and computations follow the usual patterns found in bidirectional type
systems~\cite{ChlipalaPH05,DunfieldK16}.
We show several standard-looking rules, \textrm{sub}, \textrm{lam},
\textrm{app} and \textrm{annot}.
In particular, the annotation form of \lambdaVMF, $\ottnt{e}  \mathrel{ \texttt{?:} }  a^\textsf{e}$, which
asserts the annotation $a^\textsf{e}$ correctly describes the program
$\ottnt{e}$, plays the role of type ascription in the bidirectional rules;
the \textrm{annot} rule says that terms are checked against their type
annotations, and these annotated terms synthesize the annotation type.
Because it has no dynamic semantics, \lambdaVMF uses OVV to prove and
discharge this form earlier by rewriting it to $\ottnt{e}$ sometime before
evaluation; if this rewrite fails, then the program terminates (by
failing) early, as a result of OVV failing, not execution.

As is customary in bidirectional systems, type annotations mediate
between synthesizing and checking.  This provides one the ability to
place checking-only terms (such as lambda abstractions) in positions
that require the sub-term synthesize a type (such as the abstraction
position of an application).  See typing rules \textrm{lam} and
\textrm{app}, respectively, to see the details; both are standard.
Finally, as is customary, type \emph{subsumption} allows less specific
types to check against terms that synthesize more specific types.
Rule \textrm{sub} uses a definition of type \emph{consistency}
(written $ \ottnt{C}  \approx_{\textsf{?} }  \ottnt{D} $), which behaves like type equality, modulo
the uncertain type~$ \textsf{?} $, which is consistent with all other types.
This notion of consistency is standard in some gradual typing
literature~\cite{siek+2006:gradual-typing}.

\paragraph{Gradual typing for dictionary projection.}
The typing rules for uncertain and certain projection differ in what
is known about the record and field values, and illustrate a form of
gradual typing.
In the uncertain case, rule~\textrm{p?} synthesizes return
type~$ \textsf{?} $, since nothing is known about the dictionary of values
being projected.
By contrast, in the certain case, the dictionary type is known to
rule~\textrm{p!}, and this dictionary maps the given field value to a
corresponding field type.
In this case, the soundness of the type system means that the
projection must succeed in all possible future program states, and
moreover, that the projected value has the given type.

\paragraph{Typing the \textsf{libDb} operations.}
The rule~\textrm{openDb?} is uncertain and has no certain counterpart:
The type of the database is not known until after the operation
completes, just before execution resumes with its continuation; before
then, the database could hold any type, so the rule types the returned
database as~$\ottkw{Db} \, \textsf{?}$.

Following similar reasoning, since filtering and joining databases
occur \emph{after} a database is loaded, it is possible to type these
operations in both uncertain and certain modes.
The rule~\textrm{filterDb?} says that filtering a database of
uncertain values leads to another database of values with an uncertain
type; since it merely assumes the type of the database is~$ \textsf{?} $, it
does not prove that the predicate will not ``go wrong'', e.g., by
projecting the wrong field from its argument.
By contrast, the rule~\textrm{filterDb!} says that filtering a
database of known type using a predicate that checks against this type
leads to a database with the same known type; in this case, the
soundness of the type system means that the predicate must always
succeed.

Similarly, the rules for \textrm{joinDb?} and \textrm{joinDb!} follow
the pattern set above: the uncertain rule assumes nothing about the
argument values, beyond the arguments actually consisting of
databases.  The certain rule assumes that the database arguments'
types are fully known, that the chosen field values are mapped in
these types, and that the chosen fields share a common type (we want
to compare values of this field for equality to perform the join).

\paragraph{Gradual Typing, Progressively via OVV.}
\figref{time2d-typing} illustrates using the typing rules of
\figref{type-terms} to perform progressive typing our four-line
motivating example~(\figref{full-example-code}).
The vertical and horizontal dimensions of the table list each of the
four lines; the vertical axis represents concrete execution, and the
horizontal axis represents typing the four right-hand-sides of the
program's four let-bindings, and in particular, for
\lstinline{filterDb} (on Line 2) and \lstinline{joinDb} (on Line 4),
the table indicates whether the operation was typed in the uncertain
or certain modality.
As execution progresses (vertically) over Lines 1 and 3, the
typing of the program continuation's becomes more certain: After Line
1 executes, Line 2 types in the certain modality ($ \texttt{!} $) instead of
the uncertain modality ($ \textsf{?} $); similarly, after Line 3, Lines 4
onward type using the certain modality.
As the final column shows, the certainty of code using these tables in
the remainder of the program (Line 5 onwards) increases after each of the two calls to
\lstinline{openDb}.  Between the two calls, some information is known
(relating to the first two tables defined on Lines 1 and 2), but some
information is still missing (relating to the two tables defined in Lines 3
and 4).

Our current Rust-based prototype of \lambdaVMF is powerful enough to
express this example, including the progressive typing discussed
above.
%
In \secref{discuss}, we discuss the potential to use incremental
computation in the context of such progressive typing; the goal is to
improve performance by exploiting the redundancy of re-typing the
program's continuation.

\section{Discussion}
\label{sec:discussion}
\label{sec:discuss}

In this section, we discuss future challenges and directions for the
vision of OVV presented in this paper.
Specifically, we discuss the design of the meta-level programming
language, and its use in expressing \emph{progressive verification} and
\emph{regressive validation}.

\paragraph{Incremental Computation for Progressive Verification.}
In the motivating example from \secref{overview}, the \code{chk_stk}
calls performed by \code{openDb} use progressive typing to check their
continuations. In fact, these two continuations are related: The
earlier version lacks type information about the table loaded in line
3, whereas the later version has access to this type information.
%
Progressive typing could exploit this incremental
relationship to avoid re-computing all of the typing facts about the
program state that have \emph{not changed}.

\newsavebox{\SBoxRegressive}
\sbox{\SBoxRegressive}{\includegraphics[width=0.45\linewidth]{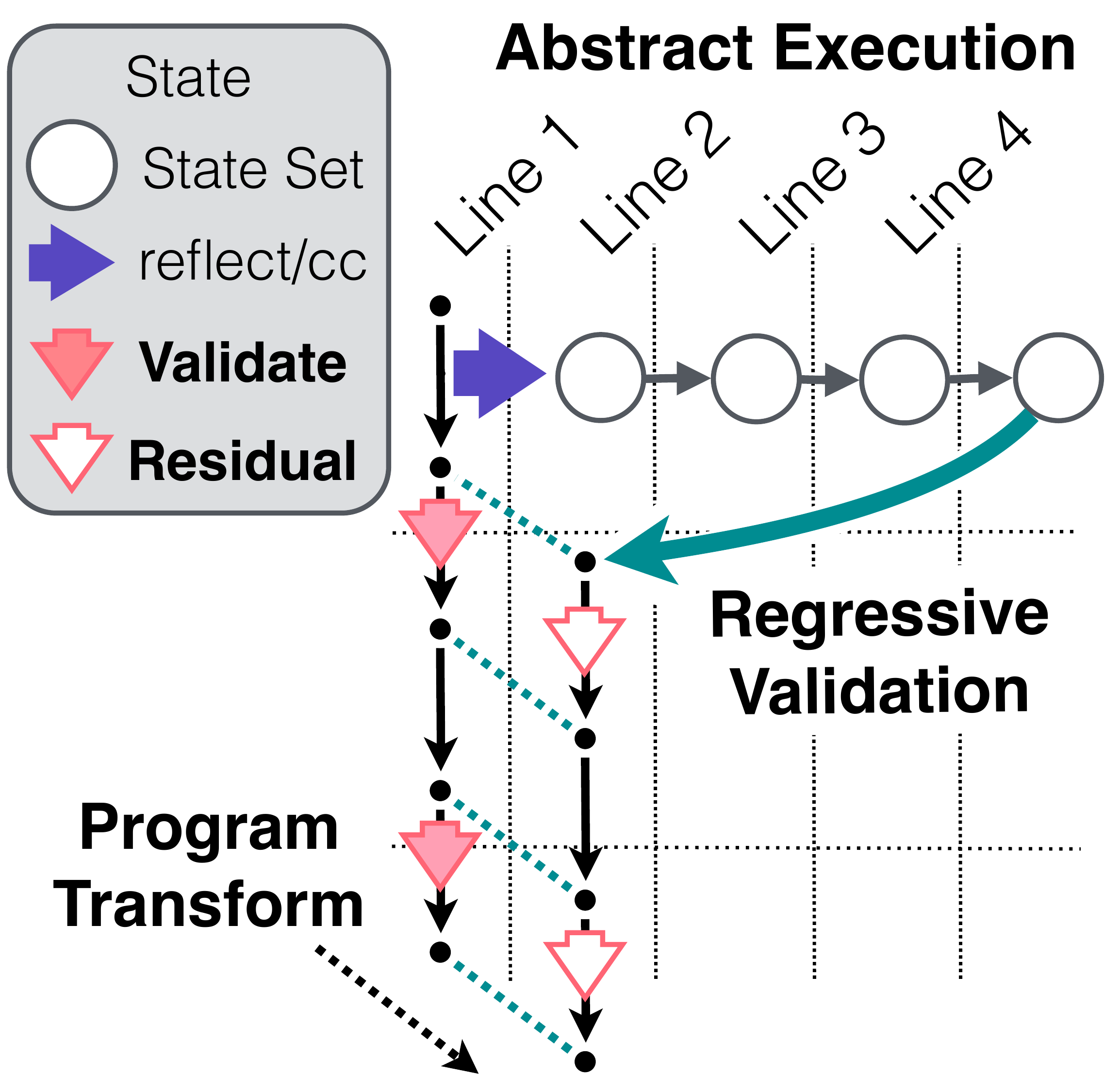}}
\newlength{\RegressiveBoxHeight}
\settoheight{\RegressiveBoxHeight}{\usebox{\SBoxRegressive}}

\begin{figure}
  \subfloat[Progressive verification]{\label{fig:progressive}
    \includegraphics[width=0.45\linewidth]{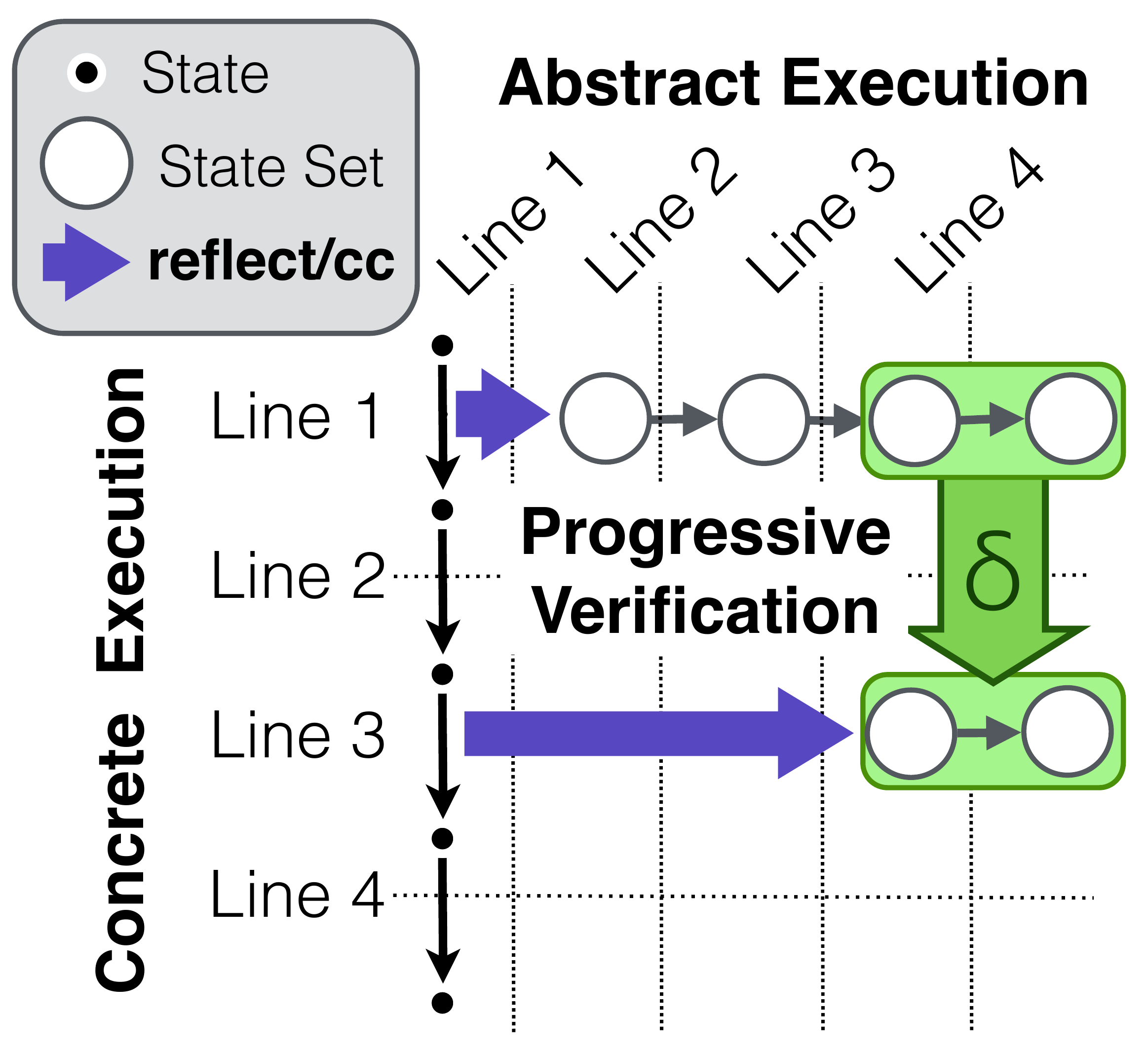}
  }
  \hfill
  \subfloat[Regressive validation]{\label{fig:regressive}
    \usebox{\SBoxRegressive}
  }
\caption{Progressive verification and regressive validation are complementary
ways of seeing the interplay between \ForallAnalysis{} and \ExistsAnalysis{}
enabled by online verification-validation.}
\label{fig:progressive-regressive}
\end{figure}

Progressive typing is a specific instance of progressive verification.
Pictorially, progressive verification relates distinct abstract
executions (shown horizontally in \figref{progressive}), by exploiting their similarity (a
small change, depicted as~$\delta$, extending vertically).
The task of a progressive verifier is a \ForallAnalysis, just like a
classical static verifier, which attempts to prove that \emph{all}
executions to an assertion satisfy a particular safety property.


\paragraph{More Aggressive Regressive Validation.}
After type checking the continuation for line 1, \emph{regressive validation}
consists of eliminating dynamic checks within the library calls of
lines 2 and 4 (\code{filterDb} and \code{joinDb}, respectively).
In particular, if the continuation type-checks under the
partially-known type information, the known type information can be
used to elide run-time \ExistsAnalysis checks that concern the authors table, including the
\lstinline{assert}s for the projections of \code{author.name}.
Pictorially, we think of regressive validation as introducing a third
dimension that consists of all possible outcomes of a program
transformation on the object program (shown in \figref{regressive}).
%
%
After performing an online verification, the meta layer transforms the
continuation, either eliding certain downstream validation checks
(labeled~\textbf{validate}), or introducing residual checks that
reduce the original checks' complexity (labeled~\textbf{residual}).
As illustrated in \secref{typing}, progressive typing can, before
executing the operation, eliminate residual checks by rewriting
uncertain operations to certain operations (which require no run-time
type checks).
%
More aggressive verification techniques can hope to
regress even more aggressive validation checks to simpler forms.
For instance, global heap-based properties present an interesting
challenge.


\paragraph{Implicitly-Incremental Meta-Level Computation.}
In sum, the example~\code{chk_stk} above encodes a theory about
\emph{online typing}, along with a mechanism for using this type
information to optimize the dynamic run-time checks that would
otherwise be used.
%
The meta layer should have an in-built ability to
implicitly express progressive verification as ordinary verification,
so that the system, not the programmer, takes into account execution
environment changes across these progressive stages.
Further, when the meta layer uses this progressive verification to
enable regressive validation in the future execution down stream, the
system, not the programmer, accounts for these changes when doing
future stages of progressive verification.
In other words,
the VM that runs the meta language and
object language should have an in-built ability to express interaction
among the levels in terms of implicitly-incremental computation.
In \secref{related-work}, we discuss the challenges that OVV poses to
work on general-purpose incremental computation.

\section{Related Work}
\label{sec:related-work}

This section supplements the related work in \secref{overview}.

\paragraph{General-Purpose Incremental Computation.}
\secref{discuss} proposes an implicitly incremental meta-level
language for \lambdaVMF, which challenges current research on
(general-purpose, programming language-based) incremental computation
(IC).
Consider the desired incremental behavior of \code{chk_stk} in the
motivating example, where it occurs after lines 1 and 3, when the
program states are similar, but not identical.
In particular, both continuations include the AST of the call in line
4, and onward, which~\code{chk_stk} will process in \emph{both}
verification stages.
The central challenge is reusing the redundant work performed
by~\code{chk_stk}, despite the fact that the AST and
store typing~\emph{are not equal to that in the prior stage}, which
creates challenges for incremental computing via \emph{memoization},
a key implementation mechanism used across many specific IC approaches.

To understand why these ``small'' changes are challenging for typical memoization,
consider the structural recursion of
the \code{chk_stk}~function from \figref{chkCont}.
One approach to memoization identifies each saved invocation by the
entire store typing~\code{st} and entire stack~\code{stk}, including
all of their recursive sub-structure, e.g., via
\emph{hash-consing}~\cite{filliatre2006type}.
This approach is commonly taken by past work on incremental
computation~\cite{PughTe89,PughThesis,PughTe89,Guo2011,Adapton2014,Bhatotia2011,Bhatotia2015,ErdwegBKKM15},
however, it is brittle and overly sensitive to small changes, since
they alter the identity of the whole recursive structure.
Recent work addresses this shortcoming by introducing \emph{unique
  names} that are special to incremental
computing~\cite{HammerDHLFHH15}.
This naming mechanism can overcome the challenges outlined above
for~\code{chk_stk}, since the presence of names isolates changed
components of the store typing, stack and local environments.

However, several key challenges remain before these techniques can fully realize
OVV: We want to use these names \emph{correctly} (to avoid unsound
incremental results), use them \emph{efficiently} (to isolate changes
and avoid sub-redundant computations) and use them \emph{implicitly}
(so that the meta-level programs look like ML).
Further, we may want to control how fine-grained the IC techniques
track program dependencies, to reduce constant-factor overhead.
%

\paragraph{Reflective Towers of Interpreters.}
As proposed in \secref{discuss}, future work on \lambdaVMF should
permit library extension authors to write meta programs and object
programs in an integrated way.
Fortunately, many researchers have proposed designs that allow
interesting interplay between the interpreter's viewpoint (where the
meta-level program runs) and the program being interpreted (where the
object program runs).
Conceptually, this work begins with 3-LISP~\cite{Smith84,RivieresS84},
which gives the programmer access to an \emph{infinite
  tower} of (so-called \emph{meta-circular}) interpreters, allowing
them to redefine the language from within the language.
Following (theoretical) work on 3-LISP, researchers give various
approaches that attack practical concerns in how to express and
implement reflective towers in simpler terms; these efforts are named
after various hair colors: Brown~\cite{FriedmanW84, WandF88},
Blond~\cite{DanvyM88} and most recently, Black~\cite{AsaiMY96,Asai11}.

Compared to the impressive and mind-bending work on metacircular
interpreters, the vision for \lambdaVMF is more modest: Two levels
suffice to perform OVV.  Having said that, if meta-level programmers
want to verify their meta-level programs as object programs (to
``bootstrap'' a typed meta level), the work mentioned above will
likely provide further insights.

\paragraph{Program Analysis for Dynamic Languages.}

The ultimate aim of online verification-validation is offer ``strong checking''
in an extensible, dynamic language environment. And thus we seek to build on the
substantial amount of research activity on program analysis for dynamic
languages. Since by definition, dynamic languages lack a built-in static typing
discipline, much of the static verification work focuses on either retrofitting
rich typing or specification
disciplines~\cite{DBLP:conf/oopsla/ChughHJ12,chugh+2012:nested-refinements:,DBLP:conf/popl/GardnerMS12}
or applying whole-program flow analysis for inferring and checking type
properties (e.g., for
JavaScript~\cite{DBLP:conf/sas/JensenMT09,DBLP:conf/sigsoft/KashyapDKWGSWH14,DBLP:conf/sigsoft/BaeCLR14} or
for Ruby~\cite{DBLP:conf/kbse/AnCF09}).

The dynamic language features that make widely-used libraries like jQuery
possible also make retrofitting static techniques incredibly
challenging~\cite{DBLP:conf/ecoop/LernerELK13,DBLP:conf/pldi/SchaferSDT13,DBLP:conf/ecoop/SridharanDCST12,DBLP:conf/oopsla/AndreasenM14}.
Much of this work focuses on finding the right kinds of context-sensitivity to
try to more precisely resolve the flow of values to dynamic features like
dynamic property read in a static
analysis~\cite{DBLP:conf/ecoop/SridharanDCST12,DBLP:conf/oopsla/AndreasenM14,DBLP:conf/ecoop/ParkR15}
or to determine when dynamically-observed information is sufficient to apply in
a static verification~\cite{DBLP:conf/pldi/SchaferSDT13}. We expect such
techniques to be not only applicable and useful but strengthened in an OVV
context. In the end, static techniques in a phasic setting are limited by what
is indeed available statically, and dynamic techniques are limited by what can
be observed in testing runs. As exhibited in \secref{overview}, the vision of
OVV enables these techniques to be strengthened with a flexible interleaving of
``static'' and ``dynamic'' analysis (i.e., \ForallAnalysis and \ExistsAnalysis).

\section{Conclusion}
\label{sec:conclusion}

This paper presents a vision for online verification-validation (OVV),
an approach to ease the tension between extensibility (of dynamic
languages) and safety (of static languages).
The key insight of OVV is that analysis in a VM can be
\emph{phaseless}, allowing analyses to run progressively on the object
program by pausing execution, reflecting on the current continuation,
and transforming the continuation to replace \emph{uncertain}
(\textsf{?})  operations with \emph{certain} (\texttt{!}) ones.

In this paper, we formalize an approach for OVV as a language
semantics and Rust-based implementation called \lambdaVMF.
We explore a proof-of-concept instantiation of OVV by defining a
gradual type system for dynamic field projection and databases with
dynamic schemas, and we observed that the result is a progressive type
checker.


\clearpage
\bibliographystyle{plainnat}
\bibliography{conference.short,bec.short,hammer-thesis1.short,hammer-thesis2.short,adapton.short,dvanhorn-local,dvanhorn,reflective.short,gradual,bidir.short}

\end{document}